# Probing up-conversion electroluminescence of decoupled porphyrin molecules in a plasmonic nanocavity

*Li-Qing Zheng[1]*[†], Fábio J. R. Costa[1,3], Abhishek Grewal[1], Ruonan Wang[4], Fengmin Wang[5], Wei Li[5], Anna Rosławska[1]*, Klaus Kuhnke[1]*, and Klaus Kern[1,2]*

1. Max-Planck-Institut für Festkörperforschung, 70569 Stuttgart, Germany

2. Institut de Physique, École Polytechnique Fédéral Lausanne, 1015 Lausanne, Switzerland

3. Gleb Wataghin Institute of Physics - University of Campinas - UNICAMP, Campinas 13083-859, Brazil

4. State Key Laboratory of Analytical Chemistry for Life Science, School of Chemistry and Chemical Engineering, Nanjing University, Nanjing 210023, People's Republic of China

5. Key Laboratory of Mesoscopic Chemistry, School of Chemistry and Chemical Engineering, Nanjing University, Nanjing 210023, People's Republic of China

ABSTRACT

Molecular triplet states can produce significant phosphorescence and act as a relay state for luminescence, such as in up-conversion processes. While this property makes triplet emitters interesting for organic light-emitting diodes (OLEDs), the study of their luminescence at the single molecule level in high resolution scanning tunneling microscopy (STM) is challenging. We investigate individual Pd-octaethylporphyrin (PdOEP) molecules decoupled from Ag(100) and Ag(111) by an ultrathin NaCl layer and observe singlet and triplet emission lines at visible wavelengths, only about 100 nm apart from each other. This is in stark contrast to the metal or free-base phthalocyanines, for which typically the lowest triplet transitions lie in the far red or infrared where the sensitivity of charge coupled device (CCD) detectors decrease significantly. The singlet $S_1$ state of PdOEP emits photons even when the photon energy is higher than the energy provided by a tunneling electron, in an energy up-conversion process. This mechanism requires a relay (or shelving) state in which energy is stored in the molecule for the interval between tunneling electrons. Analyzing the energy levels of different molecular states ($S_1$, $D_0$, and $T_1$ states) and fitting the current dependencies of $S_1$ under up-conversion electroluminescence (UCEL) condition for $S_1$ and $T_1$ emission, we verify the validity of a triplet-mediated up-conversion model.

INTRODUCTION

The study of triplet state excitation and relaxation pathways of a molecule is highly desired for improving the energy efficiency of organic light-emitting diodes (OLEDs)[1, 2, 3]. Ever since the first report of phosphorescence in OLEDs, an important strategy for lowering excitation energy in optoelectronic devices is the development of highly efficient triplet emitters due to their general superiority that results from lifting the requirement of strictly antiparallel spin injection. Over the years, this approach evolved towards synthesizing systems in which the energy stored in the triplet state is used to generate efficient fluorescence, as done in the 3rd and 4th generation devices.[4] In that respect, probing the interplay between the singlet and triplet emission is an efficient strategy to study and understand the processes playing a key role in triplet relaxation, and as such requiring efficient triplet emitters. Metal porphyrins containing heavy metal elements such as Pd-octaethylporphyrin (PdOEP) are notorious for intense phosphorescence in solution[5-7] and in a solid matrix[8], due to the strong spin-orbit coupling at the center Pd atom. PdOEP is frequently utilized as an effective triplet emitter in ensemble electroluminescence systems, including OLEDs.[9] However, developing a more profound understanding of the fundamental exciton formation mechanisms in thin film devices is challenging due to the intermolecular coupling between the molecules. This issue can be addressed by studying the electroluminescence of PdOEP at the single-molecule level, which may provide key insights into developing more efficient emitters for OLEDs.

Detection of light emitted from the tunnel junction of a scanning tunneling microscope (STM) enables the study of single-molecule emitters[10-22]. An STM provides the necessary spatial resolution down to the atomic level and allows for the precise control of the molecule's local environment in combination with atomic-scale manipulation techniques. STM-induced luminescence (STML) with submolecular resolution has been successfully applied to the study of the fluorescence of individual neutral and charged molecules over the past years[13, 16, 20, 23-27]. These studies resolved molecular electronic excitation and radiative relaxation pathways, and have to take into account both singlet and triplet states.[20, 26, 28, 29]

Electroluminescence with a photon energy higher than the energy of a single injected or extracted electron is referred to as up-conversion electroluminescence (UCEL). It has received much attention in several single-molecule STML studies.[26, 29-32] Possible UCEL mechanisms comprise triplet-triplet annihilation[33, 34], thermally assisted activation[35-37], Auger processes[38], and triplet-assisted processes[26, 29]. Earlier studies on individual metal phthalocyanine molecules[26, 29, 32] postulated that UCEL from the lowest singlet $S_1$ state involves a previous excitation to the lower-lying triplet $T_1$ state, because the $S_1$ state cannot be energetically accessed by a single tunneling process from the STM tip. In these studies, however, the singlet emission was not correlated with the triplet luminescence, since detecting phosphorescence in STML is extremely challenging, for two reasons. First, the $T_1$ emission energy of metal or free-base phthalocyanines typically lies in the far red or infrared where the sensitivity of photon detectors decreases significantly.[28] Second, a long-lived $T_1$ state is likely to decay to a charged doublet state before light emission can occur.[20, 24]

Here, we use single PdOEP molecules decoupled from Ag surfaces by 3 monolayers (MLs) NaCl and simultaneously monitor $S_1$ singlet emission (fluorescence) and $T_1$ triplet emission (phosphorescence) excited by the current of an STM (Figure 1a). The relation between these two emission lines opens a path to analyze UCEL and triplet relaxation in an unprecedented way. Intriguingly, we find that the intensity ratio of $S_1$ and $T_1$ emission lines stays approximately constant when lowering the electron energy across the transition to the UCEL range. We examine the voltage- and current-dependent electroluminescence and perform power-law analyses of intensity versus voltage and tunneling current. By analyzing the current dependencies of $S_1$ and $T_1$ emission intensities, we verify an up-conversion model, in which the $T_1$ state acts as a relay state for the UCEL of $S_1$.

## RESULTS AND DISCUSSION

PdOEP molecules are sublimed to provide a coverage of only a few percent of a monolayer on a pre-cooled (~100 K) Ag(111) substrate that was previously covered by NaCl multilayers but leaving areas of the clean substrate, where the tip can be prepared. The sample was then transferred to the STM main chamber and studied in a home-built STM setup operated at T = 4.2 K.[39] For details, see the experimental section in the Supporting Information (SI). The experimental setup is sketched in Fig.1a. In contrast to metal phthalocyanines, which adsorb individually on NaCl terraces, PdOEP molecules exhibit a high mobility on NaCl layers even at low temperature (100 K). This is due to their non-planar structure caused by the steric hindrance of the ethyl groups, which weakens the van der Waals interaction between a molecule and NaCl, as discussed in our previous work.[40] Consequently, when evaporated onto a substrate held at ~100 K, PdOEP molecules preferentially attach to other molecules or to step edges on the NaCl surface (inset of Figure 1c), demonstrating their mobility upon contact with NaCl. In order to obtain stable experimental conditions, STM images and emission spectra have to be acquired at low tunneling current (< 50 pA). Unless otherwise stated, all measurements in this study have been performed on molecules adsorbed on 3ML of NaCl on Ag(111) to provide both sufficient decoupling from the metal and electron extraction from the positive ion resonance (PIR) of PdOEP as close as possible to the Fermi energy[41]. Under low tunneling current conditions, we are able to study luminescence in the UCEL range of PdOEP fluorescence, which is the main goal of this study and will be discussed in detail below.

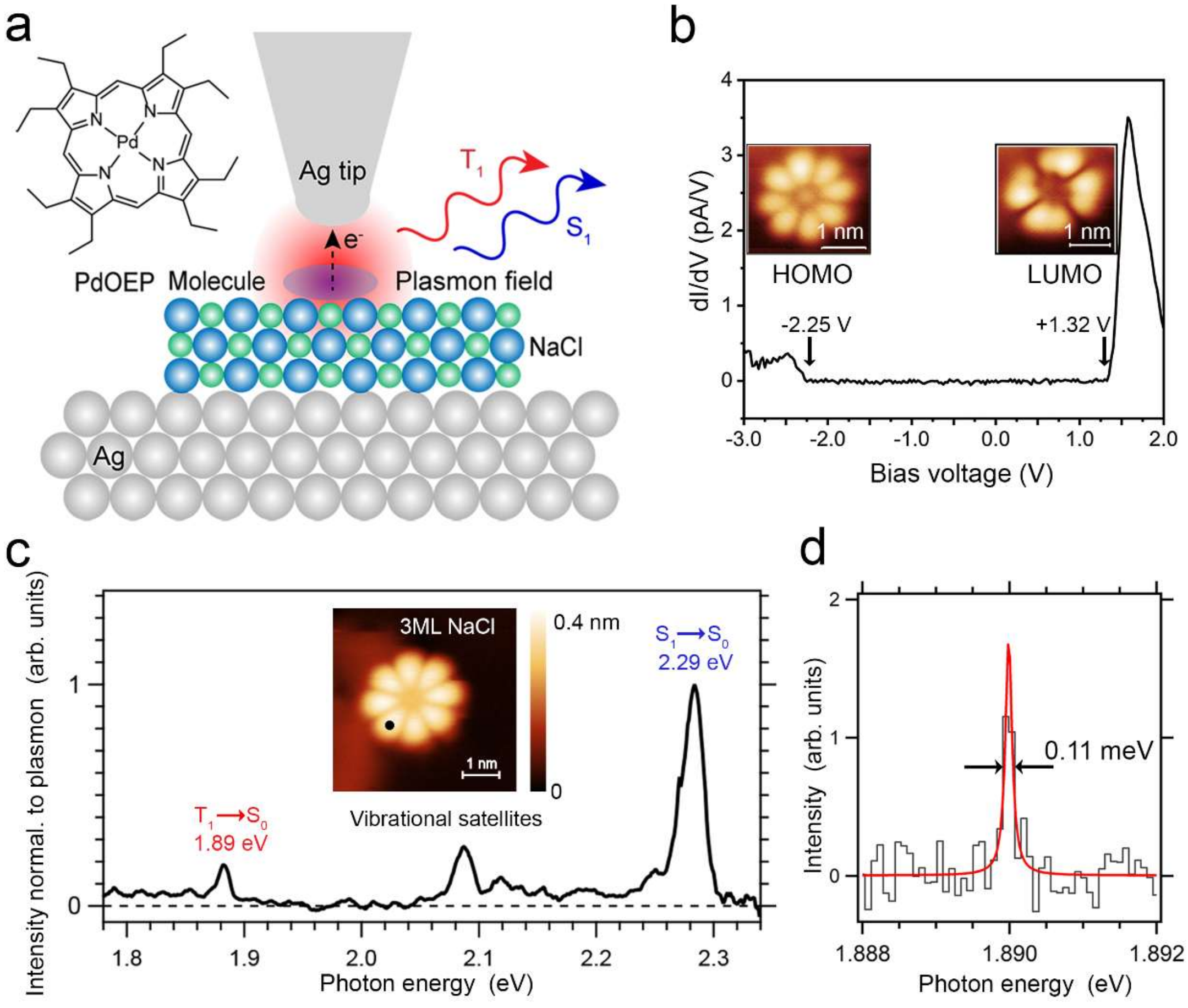

**Figure 1.** a. Schematic of the excitonic emission ($S_1$ and $T_1$ emissions) of a PdOEP molecule decoupled from an Ag substrate by 3ML NaCl under the excitation of tunneling electrons. b. Typical differential conductance (dI/dV) spectrum of a PdOEP molecule on 3ML NaCl / Ag(111). Insets are the HOMO and LUMO images of the PdOEP molecule acquired at -2.5 V and +1.7 V, respectively, on 2ML NaCl / Ag(111) (I = 1 pA). c. Typical STM electroluminescence spectrum of a PdOEP molecule on 3ML NaCl / Ag(111). Measurement conditions: I = 30 pA, V = −2.5 V, t = 360 s, 150 lines/mm grating. The inset shows an STM image of the molecule attached to a NaCl step with the tip position indicated for the emission measurement (tunneling conditions: I = 1 pA,

V = −2.5 V). The spectrum is normalized to the plasmonic emission spectrum of the tip (see Figure S1). d. Emission spectrum (raw data) of $T_1$ acquired with a 1200 line/mm grating (gray curve). The red line is a Lorentzian fit providing the indicated FWHM of the corresponding 0-0 peak.

Figure 1b shows a typical differential conductance (dI/dV) spectrum of a PdOEP molecule. The onset of the PIR and the negative ion resonance (NIR) are found at -2.25 V and +1.32 V, respectively. The insets show STM orbital images measured at -2.5 V and +1.7 V, consistent with calculated images of HOMO and LUMO, respectively.[39] First, we focus on the luminescence properties of PdOEP under electronic excitation by the tunnel current of the STM. Figure 1c shows a plasmon-normalized luminescence spectrum of a PdOEP molecule on 3ML NaCl/Ag(111) obtained at a sample bias of -2.5 V with the tip placed on one of the HOMO lobes (black dot in the inset of Figure 1c). For the raw data and the normalization procedure see Figure S1. Two distinct emission peaks at 541 nm (2.290 eV) and 656 nm (1.890 eV) are assigned to the singlet ($S_1$) and triplet ($T_1$) emission of PdOEP, respectively. The observed wavelengths are in very good agreement with photoluminescence spectra of PdOEP in solution[42] (exhibiting only a ~25 meV blue-shift) and in Shpolskii matrix at 77 K[5](with a ~42 meV blue-shift). Note that, in STML, the emission properties of $S_1$ and $T_1$ are also sensitive to the molecular local environment, such as the plasmonic enhancement, layer thickness of NaCl (Figure S2) and the adsorption orientation of PdOEP (Figure S3), whereas the emission wavelengths that we observe on Ag(100) and Ag(111) substrates agree within ~4 meV or less (Figure S4). For PdOEP on 2ML NaCl on Ag(100), the emission peak of $S_1$ at 2.290 eV is observed, while the $T_1$ emission peak is missing (Figure S2). The lack of the $T_1$ emission peak can be attributed to the low plasmon enhancement at the energy of the $T_1$ emission and efficient non-radiative $T_1$ quenching channels due to the thinner buffer layer. Notably, compared with the differential conductance spectrum of PdOEP on 3ML NaCl (Figure 1b), there is no noticeable shift in the molecular resonances for PdOEP on 2ML NaCl. Moreover, no peak shift is observed for the $S_1$ emission of PdOEP on 2ML NaCl (also close to the edge of another salt layer, Figure S2), which indicates that neither the thickness of NaCl nor the nearby salt edges strongly affect the emission energy of $S_1$. While $T_1$ emission is dominant in solution due to the efficient intersystem crossing from $S_1$ to $T_1$, the $S_1$ emission is clearly dominant in the STM electroluminescence spectrum. This indicates that the tip-enhanced emission mechanism of PdOEP in the STM junction favors the $S_1$ emission over the $T_1$ emission. The STM plasmonic nanocavity increases the radiative decay rate of the emitter (Purcell effect) and shifts its emission energy (Lamb shift), as demonstrated by the STM electroluminescence of $H_2Pc$ on 3ML NaCl/Ag(111).[21] Notably, several peaks in the range from 2.067 to 2.255 eV are assigned to vibrational satellites of the $S_1$-$S_0$ transition. The vibrational peak at 2.095 eV is ascribed to the C=C vibrations of the pyrroles of PdOEP corresponding to the Raman peak at 1547 $cm^{-1}$ (the energy of the excitation laser for the confocal Raman measurement is 2.54 eV), corroborated by the simulated Raman spectra of PdOEP (Figure S5). The result of the simulated Raman spectrum is obtained by density functional theory (DFT) calculations.

Figure 1d shows that the $T_1$ emission peak exhibits a width of 0.11 meV only, which is limited by the instrumental resolution (Figure S6). Despite the coupling to the plasmon in the STM, the triplet emission line width in the low temperature STM luminescence spectrum is significantly less broadened than in the solution spectra. The $T_1$ line is thus more than 50 times narrower than the $S_1$ emission peak with a width of 7.0 meV (Figure S7). Using the relation $\Delta\tau \cdot \Delta E \geq \hbar$, this is consistent with a significantly longer lifetime of the triplet state than of the singlet state. However, a direct conversion to the lifetime is not possible because dephasing processes may play a significant role. For instance, it has been shown that emission peaks measured in STM-electroluminescence may be broadened by low energy vibronic transitions.[22, 32, 43] We can thus obtain only a lower limit for the lifetimes of the singlet and triplet state of 94 fs and 6 ps, respectively. The estimated value for the lower bound of the $T_1$ lifetime of PdOEP agrees extremely well with the value reported for the $T_1$ emission of PtPc.[28] The significantly larger value of the $T_1$ state lifetime agrees with the general observation, that triplet states exhibit longer lifetimes than singlet states. The integrated intensity ratio of triplet to singlet peak in Figure 1c is 1:14 after normalization to the plasmon spectrum and thus correcting for the wavelength-dependence of plasmonic enhancement and detector sensitivity. While the spin statistics of electronic excitation would suggest that $T_1$ and $S_1$ states are formed in a 3:1 ratio from the injected charges,[2] the observed deviation from that ratio suggests more complex mechanisms at play in the system. Moreover, non-radiative decay rates and tip enhancement will affect the two transitions differently and may favor the radiative decay of the singlet state in comparison to the triplet state leading to the observed low ratio.[28]

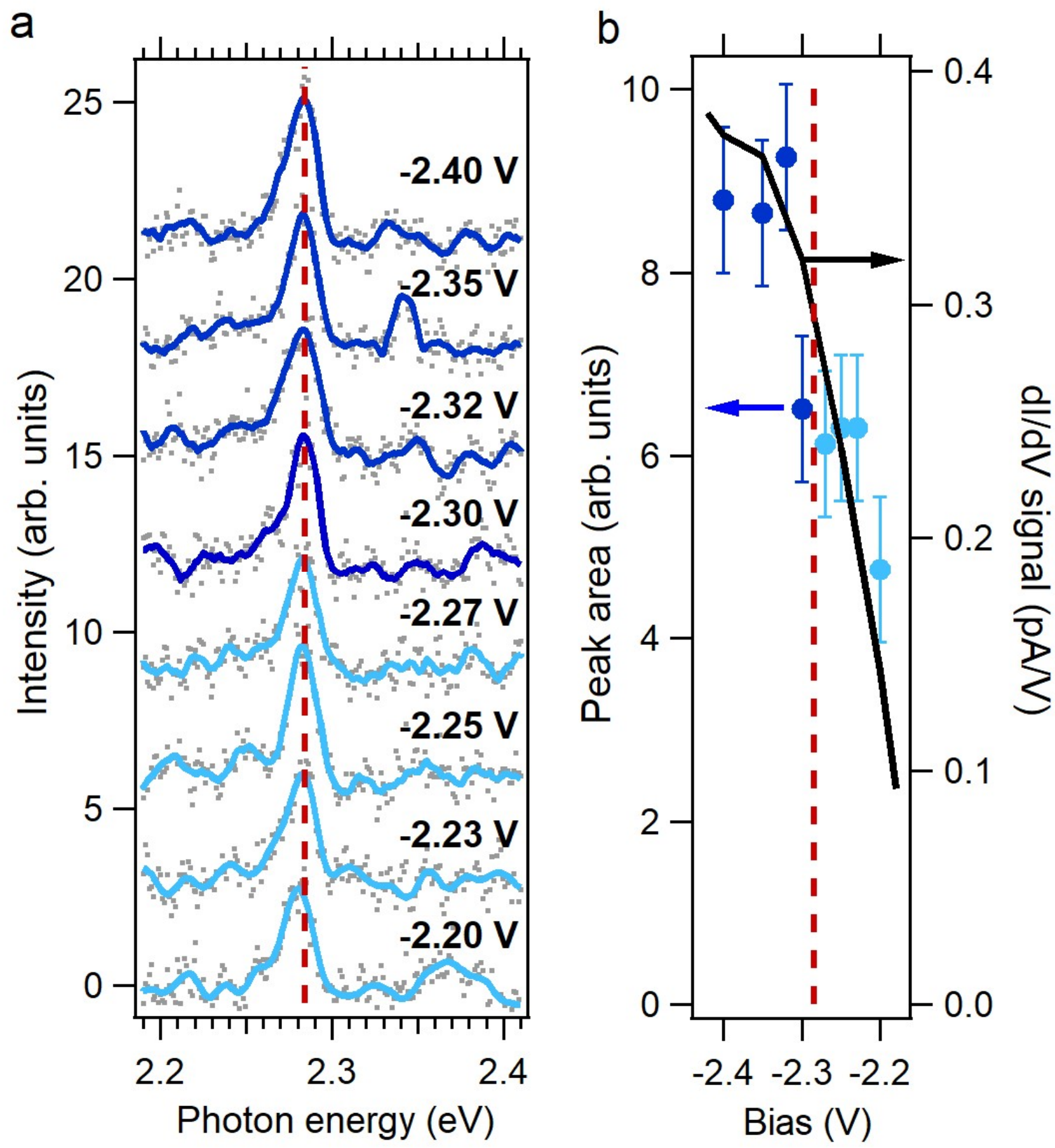

**Figure 2.** a. STM luminescence spectra (gray dots) of a PdOEP molecule ($S_1$ emission) as a function of voltage together with the Savitzky-Golay-smoothed spectra (solid lines), (I = 35 pA, acquisition time for each spectrum: 120 s). A 1200 lines/mm grating is used. The spectra are vertically shifted for clarity. The dark blue spectra correspond to the range in which single electron excitation is sufficient to induce luminescence while the cyan spectra lie in the range of up-conversion emission. b. dI/dV lock-in signal as a function of voltage (black curve) and fluorescence intensity (blue and cyan dots) of PdOEP obtained from the evaluation of (a). The vertical dashed red lines in a and b indicate the photon energy of the $S_1$ emission line.

Figure 2a displays a series of spectra in the $S_1$ emission range in which the bias voltage is tuned from -2.4 V to -2.2 V thus exploring the transition from energy-allowed excitation to UCEL, for which the electron energy decreases below the energy of the emitted photons (2.290 eV). We find that, despite a continuous reduction of intensity, the emission line remains visible in the UCEL range. In Figure 2b the evaluated intensities of the spectra are plotted as a function of voltage and are compared to the dI/dV measurement that represents the electronic density of states (DOS) of the molecule for different voltages. We remark that slight mismatch of the PIR onset energies observed between Fig. 1b and Fig. 2b can be attributed to the effect of the local environment.[44] Figure 2 thus illustrates the existence of UCEL for the studied system and indicates that the emission intensity in the transition range closely follows the molecular DOS, without cut-off at 2.29 eV. To exclude the influence of local environment shifts and tip changes on the UCEL, we show additional data across multiple PdOEP molecules and different tip conditions (see Figures S8-S10 in the Supporting Information). These results consistently show a UCEL window ranging from 40 to 90 meV. The persistence of this phenomenon across various environments demonstrates that the UCEL behavior is an intrinsic property of the system rather than a result of systematic uncertainty.

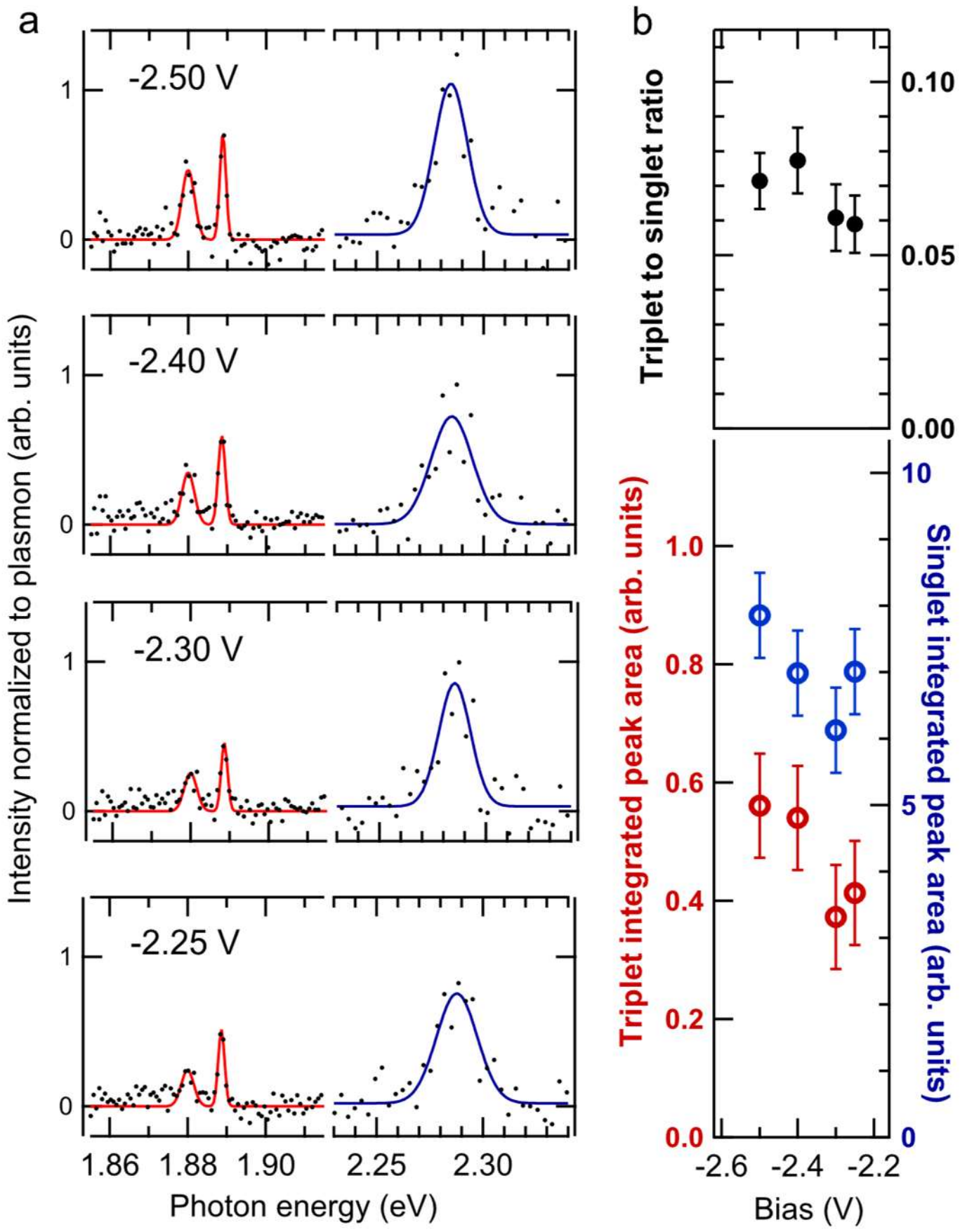

**Figure 3.** a. Two sections of STM luminescence spectra of a PdOEP molecule normalized to the plasmon intensity as a function of voltage (black dots) together with Gaussian fits to the three main peaks, representing the $S_1$ emission (solid blue line) and the $T_1$ emission (red solid line) with a satellite (which might be of vibrational origin or due to degeneracy lifting of the orthogonal electronic $S_1$ transitions; see text for discussion). For all spectra the current is I = 42.5 pA, acquisition time per spectrum from top to bottom: 450s, 600s, 600s, 750s. A 1200 lines/mm grating is used. The different acquisition times have been compensated by scaling the intensity. For the plasmon spectrum and the full range of the spectra see Figures S11 and S12, respectively. In the $S_1$ range each dot represents the three-point average (binning) of the raw data. In both spectral regions the minor ticks mark an energy difference of 10 meV. b. Lower panel: Peak area of fluorescence at 2.29 eV (blue circles) and main phosphorescence peak at 1.89 eV (red circles)

obtained from fits in (a) and plotted as a function of bias voltage. Error bars are given as twice the standard deviation (95% confidence interval) of the point-to-point scatter in the data. Upper panel: $T_1/S_1$ peak area ratio (black dots) calculated from the red and blue circles in the lower panels.

To investigate the exciton formation mechanism, we next examine spectra covering both the $S_1$ and the $T_1$ emission lines across the transition into the UCEL range. Figure 3a shows the $S_1$ and the $T_1$ emission regions of a PdOEP molecule acquired for bias voltages from -2.50 V to -2.25 V. The spectra are normalized to the plasmonic spectrum. Both fluorescence and phosphorescence emission are clearly observed at all voltages. An additional, broader peak (1.880 eV) on the low energy side of the $T_1$ emission is also observed. As the simulated Raman spectrum of PdOEP on 3ML NaCl (Figure S5b) shows no strong peak at the corresponding frequency (70 $cm^{-1}$), assigning this peak to a $T_1$ vibrational satellite is unlikely. Instead, we speculate that one of the possible origins of the second peak is the degeneracy lifting of the two $T_1$ emission modes. This lifting of degeneracy in the transition dipole moments is potentially caused by local strain from defects beneath the molecule. This explanation is supported by STML studies of ZnPc, where a Cl vacancy induced a similar degeneracy lifting of 8 meV in its optical transition.[45] While further statistical analysis of the correlation of the peak splitting magnitude with the local adsorption environments would strengthen the argument, such a measurement is technically demanding. Note that a splitting of the $S_1$ mode by a comparable energy difference would not be observable due to its natural peak width. We emphasize that we still observe the $S_1$ and $T_1$ emission at -2.25 V, which corresponds to the condition of UCEL for the $S_1$ emission (2.290 eV) of PdOEP. In the same range, the $T_1$ emission remains visible. For the full spectra shown in Figure 3, see Figure S12.

It is interesting that the ratio between $T_1$ and $S_1$ emission in Fig. 3b stays approximately constant across the transition into UCEL. This may at the first sight appear unexpected because the triplet can always be excited in a one-step process, as the energy carried by one electron is sufficient, while the singlet excitation in the UCEL range requires two electron tunneling events. However, both excitations have the same threshold voltage defined by the PIR edge. It is important to emphasize that all data in Figure 3 are taken at the same current (42.5 pA) which is the most relevant parameter determining the up-conversion efficiency from $T_1$ to $S_1$. In contrast, the energy difference between $S_1$ and $T_1$ is much smaller than the energy of $T_1$ so that no second voltage threshold is involved in the transition to UCEL. Over the examined voltage range, the $T_1/S_1$ occupation ratio and consequently the intensity ratio stays basically constant. This can be explained by the observation in previous studies[26, 32] that the two- or multi-step excitation of $S_1$ can remain dominant over a one-step excitation when the voltage crosses the energy of the emission line. Obviously, here this principle is also operative, so that even above the energy threshold the one-step excitation process is already weak and its full suppression at lower voltage does not significantly reduce the efficiency of $S_1$ excitation or may even lead to the opposite effect.[26]

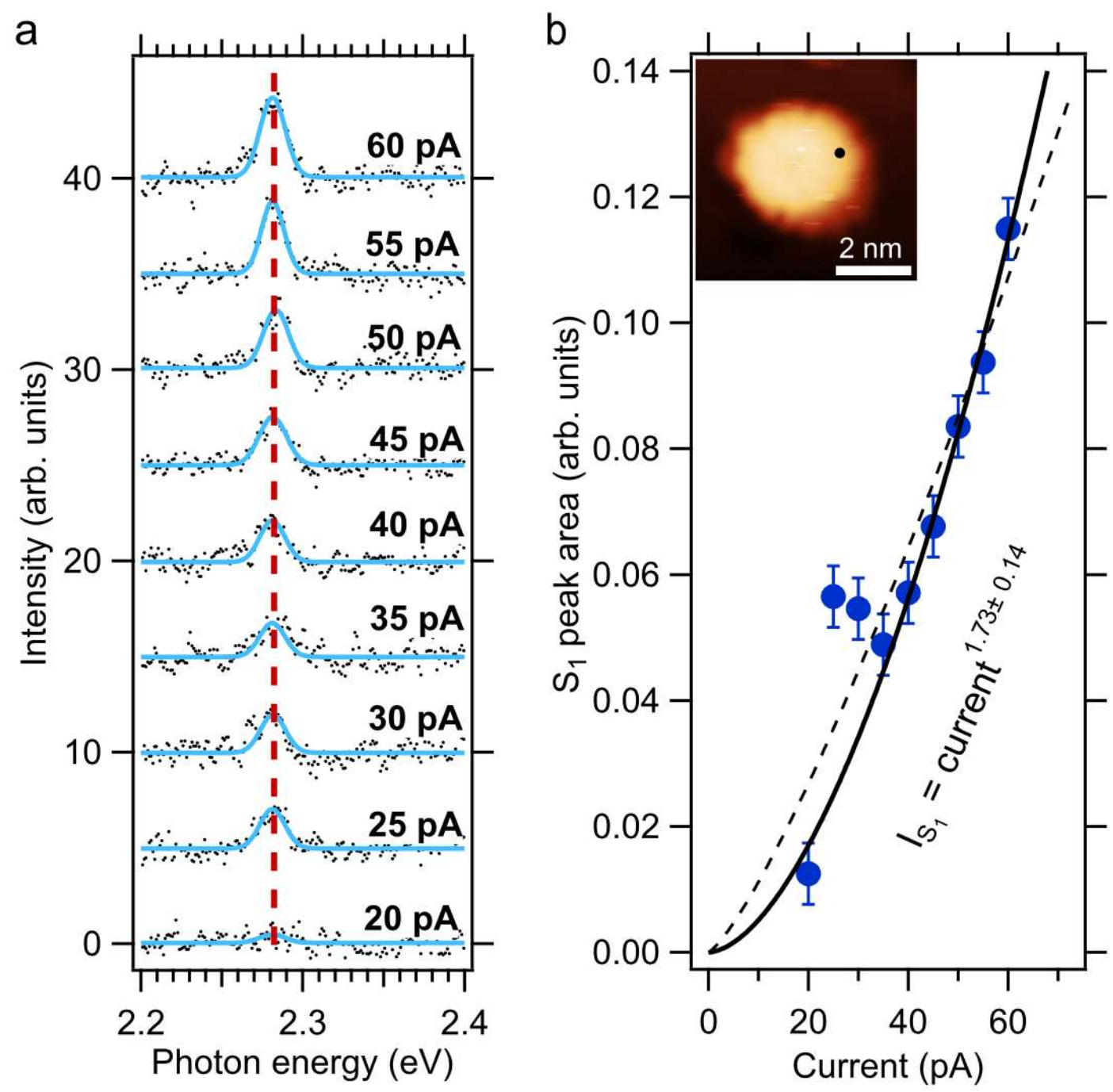

**Figure 4.** a. Current-dependent STM luminescence spectra of a PdOEP molecule (singlet emission) with Gaussian fit to the emission peak. (V = -2.25 V, acquisition time per spectrum: 120 s, current 20 pA to 60 pA as indicated at each curve). A 1200 lines/mm grating is used. The spectra are vertically offset for clarity. b. Fluorescence intensity of the PdOEP molecule as a function of current with power law fits. Blue dots: integrated peak area obtained from the fits in panel a. The dashed black curve is the best power law fit using all data points. However, the measurements at 25 pA and 30 pA appear as obvious outliers. The solid black curve is fitted with these two data points omitted so that the power law fit passes through the data points within their error bars. The inset is the STM image of the PdOEP molecules with tip position indicated (black dot).

In a further step, we now investigate the current dependence of the emission intensities of the $S_1$ line (Figure 4) and the $T_1$ line (Figure 5). Due to the low $T_1$ intensity, it was not possible to measure the current dependence still within the UCEL range of the $S_1$ emission. However, we can assume that the current dependency of $T_1$ does not change significantly between -2.25 V and -2.5 V because in contrast to $S_1$ emission, the $T_1$ emission energy does not impose a threshold value in this range which may affect the dynamic behavior. The spectrally integrated intensities are determined from Gaussian fits of the spectra in Figure 4a and 5a. Then the current dependence is fitted by power laws with two fitting parameters using a zero offset since no light is generated when no current is applied (Figure 4b and 5b). A possible baseline offset in the spectra due to noise or a plasmon background is already eliminated by the fit of the emission line. The exponent for the $S_1$ intensity is found between 1.7±0.1 (solid line) and 1.3±0.1 (dashed line), demonstrating a significant deviation from linearity as is expected for the UCEL range. A slightly larger exponent (1.8) was obtained for the current dependence measurement of the UCEL of $H_2Pc$[26], which was used to explain the multiple electron excitation pathway in the UCEL process. However, we emphasize that the exponent is not integer 2, as a pure two-step process might require. In contrast, Figure 5 shows that the exponent of the $T_1$ dependence is found at 0.8±0.3, thus indicating an almost linear behavior.

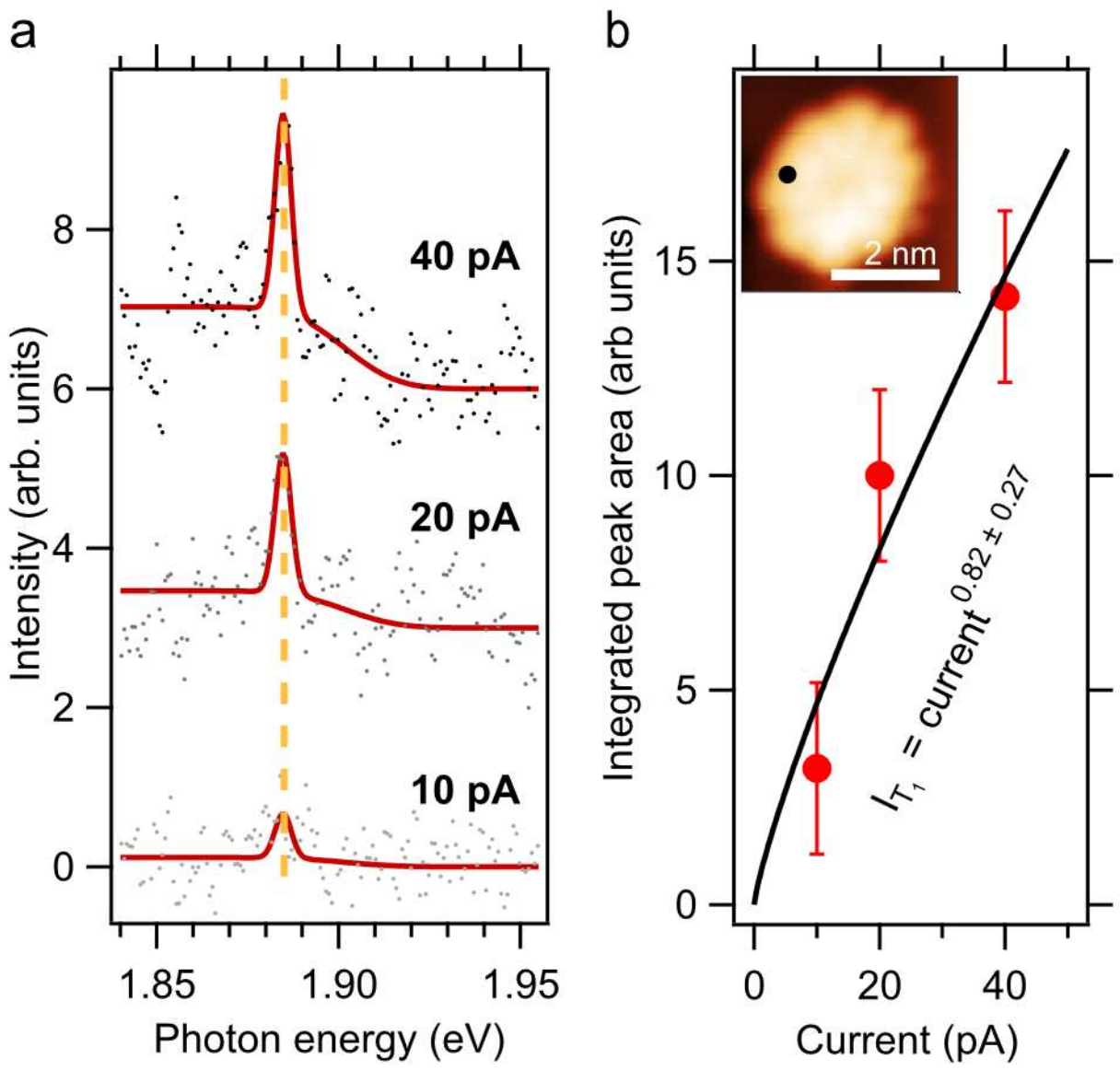

**Figure 5.** a. Current-dependent STM luminescence spectra of a PdOEP molecule (triplet emission) with Gaussian fit to the emission peak. (V = -2.5 V, acquisition time per spectrum: 120 s, I = 10 pA to 40 pA as indicated). A 1200 lines/mm grating is used. In the fit we accounted for a step appearing in the background intensity that tends to scale with the current (or emission intensity) and may be due to inelastic processes. The spectra are vertically offset for clarity. b. Phosphorescence intensity of the PdOEP molecule as a function of current with a power law fit. Red dots: Integrated peak area from the Gaussian fit. The inset is the STM image of the PdOEP molecules with tip position indicated (black dot).

With these results, we are now able to derive the fundamentals of the PdOEP excitation mechanism in STML, which we sketch as a many-body diagram in Figure 6 using the energies derived from the dI/dV and optical spectroscopy experiments.[20, 27, 46] In the voltage range explored in this study, the molecule, which initially is in the ground singlet ($S_0$) state, can be driven to the positively charged doublet state ($D_0^+$) at the PIR, marked as 1 in Figure 6, via tunneling of an electron from the molecule to the tip (black arrow). In the next step, the molecule can be neutralized by an electron tunneling from the Ag(111) substrate, which transitions the molecule back to the $S_0$ or to $T_1$ (grey arrows), indicated as 2 in Figure 6. Here, we remark that the $S_1$ state is inaccessible from $D_0^+$, as the energy of this state is too low. When the molecule is in the triplet state, a handful of transitions can occur. First, the molecule may emit in a phosphorescence process (red arrow). As an alternative, the $T_1$ state may exchange an electron with one of the electrodes, which leads to the transitions to either $D_0^+$ or $D_0^-$ states (both are charged ground states), labelled as 3 in Figure 6. We remark here that the negatively charged doublet can be reached despite applying negative bias since its energy is below the energy of $T_1$. Furthermore, as the applied bias is sufficient, it is possible to reach the $S_1$ from the $D_0^-$ state by the tunneling of the excess electron to the tip (marked as 4 in Figure 6), which we suggest as the main excitation mechanism[23] that is at play, both in the up-conversion regime and for $V < -2.29$ V. One may also consider a spin-flip process induced by the tunneling electron ($T_1 \rightarrow S_1$ transition) as an UCEL path or a similar direct $S_0 \rightarrow S_1$ transition for voltages crossing the energy of the singlet emission line. However, the observed luminescence intensity follows the local density of states profile (Figure 2 and Figure 3), no signatures of the direct $S_0 \rightarrow S_1$ excitation are observed for $V < -2.29$ V (constant slope in Figure 2b), and such processes are found to be inefficient in other systems.[28] Therefore, we consider that such mechanisms have only a minor effect in the case of PdOEP/NaCl/Ag(111). For completeness, we also include the possible $T_1 \rightarrow D_0^+$, $D_0^- \rightarrow S_0$ and $D_0^- \rightarrow T_1$ transitions in Figure 6, which act as potential non-radiative quenching mechanisms of the molecular excitation before the $S_1$ state is reached. As one notices in Figure 6, the charge-exchange transitions from the triplet state, specifically $T_1 \rightarrow D_0^+$ and $T_1 \rightarrow D_0^-$, effectively reduce its lifetime. Such dynamics were evaluated in detail for the ZnPc/NaCl/Ag(111) system and the upper bound of the $T_1$ lifetime was approximately 1 ns, decreasing with the tunneling current.[20] Assuming that for PdOEP the charge dynamics are similar, and taking the lower limit defined by the linewidth (6 ps), we conclude that the $T_1$ lifetime of PdOEP/NaCl/Ag(111) falls into the 6 ps - 1 ns range. Note that the diagram in Figure 6 has been drawn assuming a negligible voltage drop in the NaCl layer and we do not include effects related to the NaCl reorganization upon molecular charging and discharging (see Figure S13 and associated discussion for more details).[27]

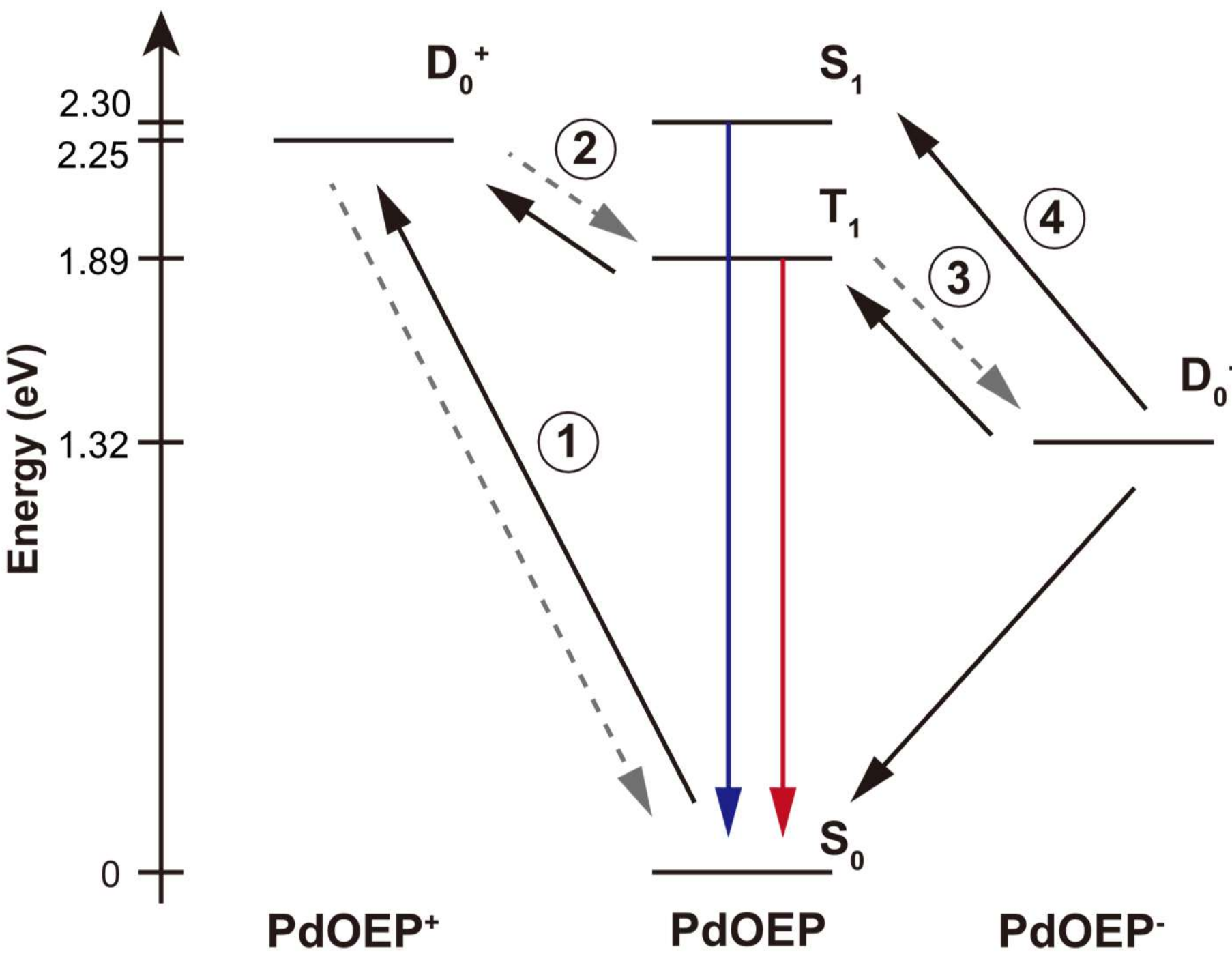

**Figure 6.** Proposed many-body diagram describing the molecular neutral ($S_0$, $S_1$, $T_1$) and charged ($D_0^+$, $D_0^-$) states. Transitions relevant for the excitation of singlet and triplet states (V < -2.2 V) are indicated by successive steps (1) to (4). The transitions drawn in black are tip-mediated, in gray (dashed) substrate-mediated. The blue and red arrows represent the singlet and triplet radiative decays, respectively. Refer to the text for a detailed discussion.

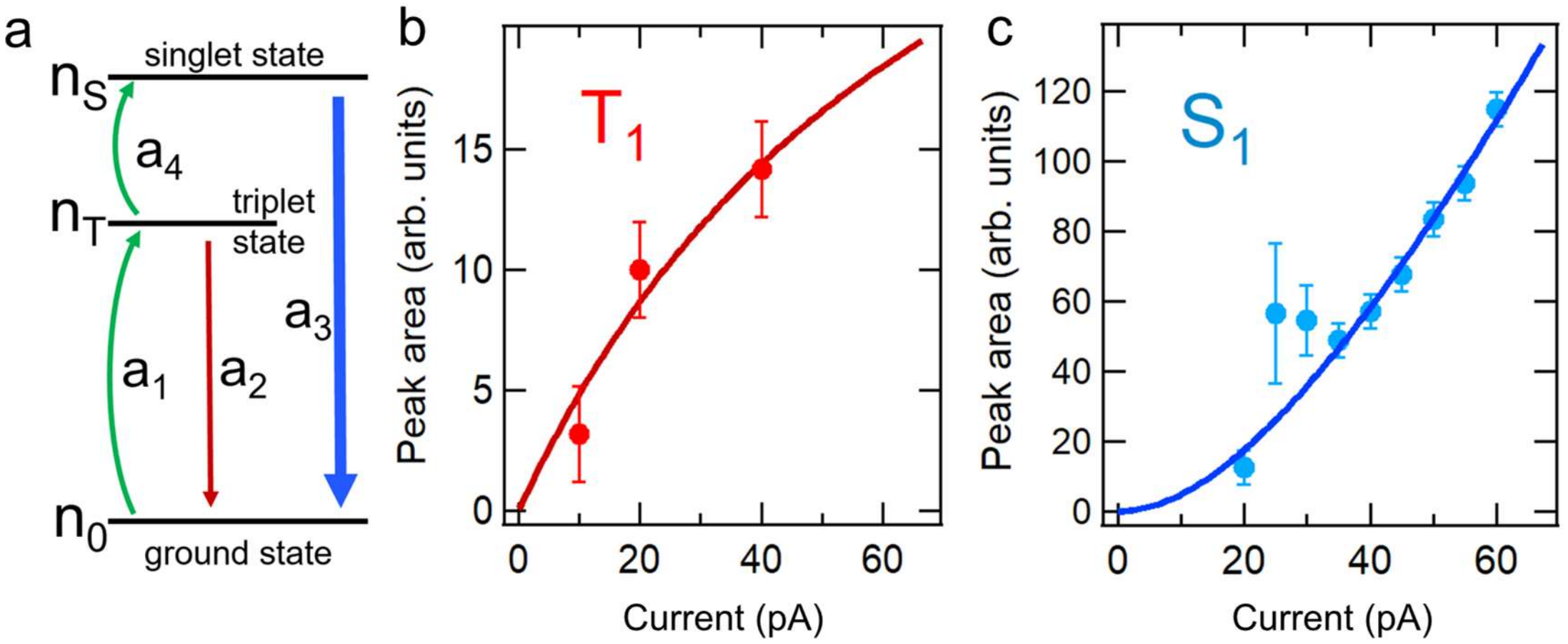

**Figure 7.** a Basic model for the up-conversion electroluminescence dynamics. The constants $a_1$, $a_2$, $a_3$, and $a_4$ indicate transition rates of the various processes. b/c. Data sets of $T_1$ and $S_1$ (dots) as a function of current from Figures 5 and 4, respectively, and a simultaneous fit of both data sets to the model (solid curves). Note that instead of excluding the two outliers in the $S_1$ data, we increased the size of their error bars with respect to Figure 4 thus relaxing their weight in the fitting procedure. For details of the model, see also Figure S15 and for details on the fit, see Figures S16, S17 and Table S1.

As a next step, we want to tackle the question, whether the obtained power laws (Figures 4 and 5) are sufficiently different from each other to confirm a basic up-conversion mechanism, or whether other mechanisms must be considered due to the fact that the $S_1$ power law exponent deviates from 2. We set up equations for the equilibrium condition for the simplified excitation scheme in Figure 7a, which are discussed in further detail in the SI. Note that the current dependencies do not provide any details of the dynamics that would make it possible to determine all individual processes that may occur between the different charge states of the molecule (Figure 6). We therefore reduce the number of included transitions to effective transition processes ($a_1$, $a_2$, $a_3$, $a_4$ in Figure 7a) that directly affect the observed intensities, thus excluding interdependent parameters. The processes $a_1$ and $a_4$ may in principle be split up into two or more steps that may render these processes non-linear in the current. However, there is no indication that such processes of even higher order in current would be required to describe the experiment. Moreover, a direct intersystem crossing transition $T_1 \rightarrow S_1$ was demonstrated to be only a minor contribution in an earlier study on Pt-Phthalocyanine under similar experimental conditions.[28] As the measurements of $S_1$ and $T_1$ expand over the same current range, we impose the strong requirement that both data series must be jointly fit with a single set of fit parameters. Figure 7b,c shows the data from Figure 4 and Figure 5, respectively, together with the obtained fits as solid curves. The result of fitting with a single parameter set suggests that the $T_1$ data actually exhibit a slightly sub-linear behavior, while the $S_1$ data shows a clear super-linear behavior. The excellent collective fit of both datasets indicates the compatibility of the data with the basic up-conversion picture. The detailed discussion in the SI elucidates that, while the exponents may readily take values differing from integer one (linearity) or integer two (quadratic dependence), the ratio of both emission intensities must stay proportional to the current itself (for all currents). This turns out to be a powerful condition providing a sharp discrimination that makes it possible to verify or reject the model. The condition implies that fitting by a single exponential is an approximation that describes the current-dependent intensities only over a limited current range. A fit of both singlet and triplet emission data with the same parameter set is thus more elucidating than the mere comparison of the exponents of separately fitted curves. The relevance of this last point is illustrated by the evaluation of the current dependence of the $S_1$ emission of a PdOEP dimer under UCEL conditions (Figure S14), which suggests the potential involvement of the triplet-triplet annihilation process. After the combination of all the data, the energy alignment in the many body diagram and the fitting of the current-dependent results confirm the proposed excitation mechanism.

CONCLUSIONS

We observe STM-induced singlet and triplet emission from a single PdOEP molecule, and monitor the emission 0.1 V into the $S_1$ up-conversion regime. This fact and the observation of the triplet ($T_1$) emission enable us to directly test the model of up-conversion from the lowest triplet to the lowest singlet state by monitoring the current-dependent and bias-dependent luminescence intensities. The observed behavior is compatible with the assumption that the triplet state acts as a shelving state that stores energy before a subsequent electron tunneling event can upconvert the molecule to the fluorescing singlet state. In earlier STML works[20, 23, 24, 26], no $T_1$ emission was observed simultaneously with the singlet emission and its role as a key intermediate state was assigned only indirectly. In contrast, we correlate the $T_1$ and $S_1$ luminescence intensities, providing direct information from the occupation of that state and its relevance for the singlet electroluminescence. These findings provide detailed insights into the UCEL and fluorescence mechanism of PdOEP at the single-molecule level, making the role of the $T_1$ state directly accessible in the electrically-driven light emission process.

## ASSOCIATED CONTENT

**Supporting Information** (see end of this document)

Details of sample preparation, confocal Raman measurements, and density functional calculations, discussion about the solution of the up-conversion model for a generic example, and supplementary figures.

## AUTHOR INFORMATION

**Corresponding Author**

* L.-Q. Zheng: lzheng@nju.edu.cn;

A. Rosławska: a.roslawska@fkf.mpg.de;

K. Kuhnke: k.kuhnke@fkf.mpg.de

**Present Addresses**

†: State Key Laboratory of Analytical Chemistry for Life Science, School of Chemistry and Chemical Engineering, Nanjing University, Nanjing 210023, People's Republic of China.

**Author Contributions**

K. Kern and K. Kuhnke supervised the project. L.-Q. Zheng designed and performed the experiments. F. J. R. Costa contributed part of the STM measurements. L.-Q. Zheng, K. Kuhnke, A. Grewal, and A. Rosławska analyzed the experimental data. R. Wang performed the Raman measurements, F. Wang and W. Li helped with the DFT simulations. The manuscript was written with contributions from all authors. All authors have given approval to the final version of the manuscript.

**Funding Sources**

This work was financially supported for L.Q.Z. by the Alexander von Humboldt Foundation and by the National Natural Science Foundation of China (No. 92478110), for F.J.R.C. by CAPES (grants 88887.517233/2020-00 and 88887.716201/2022-00), for R.W. by the National Natural Science Foundation of China (No. 92478110), and for A.R. by the Emmy Noether Programme of the Deutsche Forschungsgemeinschaft (DFG, German Research Foundation) - 534367924.

ACKNOWLEDGMENT

L.-Q. Z. thanks the financial support from the Alexander von Humboldt Foundation and the National Natural Science Foundation of China (No. 92478110). F. J. R. C acknowledges the financial support from CAPES (grants 88887.517233/2020-00 and 88887.716201/2022-00). R. W. thanks the financial support from the National Natural Science Foundation of China (No. 92478110). A. R. acknowledges funding from the Emmy Noether Programme of the Deutsche Forschungsgemeinschaft (DFG, German Research Foundation) - 534367924. We thank Olle Gunnarsson for fruitful discussions. Theoretical calculations of the Raman spectrum of PdOEP were performed at the High-Performance Computing Center (HPCC) of Nanjing University. We thank the Centre for Shared Scientific Research Facilities of Nanjing University for the Raman measurements.

ABBREVIATIONS

PdOEP, palladium octaethylporphyrin; ML, monolayer; ZnPc, zinc phthalocyanine; STM, scanning tunneling microscopy; STML, STM-induced luminescence; UCEL, up-convertion electroluminescence.

# Supporting information

## Probing up-conversion electroluminescence of isolated porphyrin molecules by scanning tunneling microscopy

Li-Qing Zheng[1*†], Fábio J. R. Costa[1,3], Abhishek Grewal[1], Ruonan Wang[4], Fengmin Wang[5], Wei Li[5], Anna Rosławska[1*], Klaus Kuhnke[1*], Klaus Kern[1,2]

1. Max-Planck-Institut für Festkörperforschung, 70569 Stuttgart, Germany
2. Institut de Physique, École Polytechnique Fédéral Lausanne, 1015 Lausanne, Switzerland
3. Gleb Wataghin Institute of Physics - University of Campinas – UNICAMP, Campinas 13083-859, Brazil
4. State Key Laboratory of Analytical Chemistry for Life Science, School of Chemistry and Chemical Engineering, Nanjing University, Nanjing 210023, People's Republic of China
5. Key Laboratory of Mesoscopic Chemistry, School of Chemistry and Chemical Engineering, Nanjing University, Nanjing 210023, People's Republic of China

**Present Addresses**

†: State Key Laboratory of Analytical Chemistry for Life Science, School of Chemistry and Chemical Engineering, Nanjing University, Nanjing 210023, People's Republic of China.

# 1. Experimental section

## 1.1 Sample preparation

All experiments were performed with a home-built low-temperature ultrahigh-vacuum STM operated at 4.2 K ($<10^{-11}$ mbar). [1] Light emitted from the tunnel junction is collimated and focused on the entrance slit of a spectrograph (Acton SP 500i or Acton SP 300i) with a charge-coupled device (CCD) camera (liquid-nitrogen-cooled or PI-MAX). Spectrograph gratings with 150 lines/mm and 1200 lines/mm were employed. Prior to use, Ag(100) and Ag(111) single crystals were cleaned by argon ion sputtering and subsequent annealing to 660 K and 670 K, respectively, for several cycles. NaCl was thermally sublimed from a Knudsen cell held at 900 K, with the Ag(100) and Ag(111) held at 300 K, to obtain a partial coverage of (100)-terminated NaCl islands. After NaCl evaporation, the substrate was annealed at 300 K for 10 min, in order to obtain 2-4 layers of defect-free NaCl islands. Finally, PdOEP molecules were thermally sublimed onto a sample held at 103 K and rapidly transferred to the low-temperature STM. Electrochemically etched Ag tips[2] used in all experiments were cleaned by argon ion sputtering to remove oxides, while electrochemically etched Au tips[3] were used without argon sputtering treatment. To clean a tip, further tip modification through voltage pulses or tip indentation was required. STM imaging and spectral measurements were performed in constant-current or constant-height mode. The text and the figure cations always specify bias voltages of the metal substrate with respect to the grounded tip and indicate the employed tunneling current. Differential conductance (dI/dV) spectra were measured using a standard lock-in technique with a bias modulation of $V_{rms}$ = 10 mV at 629 Hz.

## 1.2 Density functional theory (DFT) calculations

DFT simulations of the Raman spectra of a single PdOEP molecule and PdOEP on 3ML NaCl were carried out using Gaussian 16[4]. The DFT calculations were carried out using the PBE0[5] functional with the def2-SVP[6] basis set for all atoms except for the palladium atom, for which SDD[7] basis set was used. A scaling factor of 0.95 was used to correct the calculated Raman frequencies.[8, 9]

## 1.3 Confocal Raman measurements

The confocal Raman measurements of PdOEP were carried out using a confocal Raman spectrometer (Invia Renishaw, England) equipped with an excitation laser of 488 nm and a 100x objective of NA 0.7.

## 2. Supplementary figures

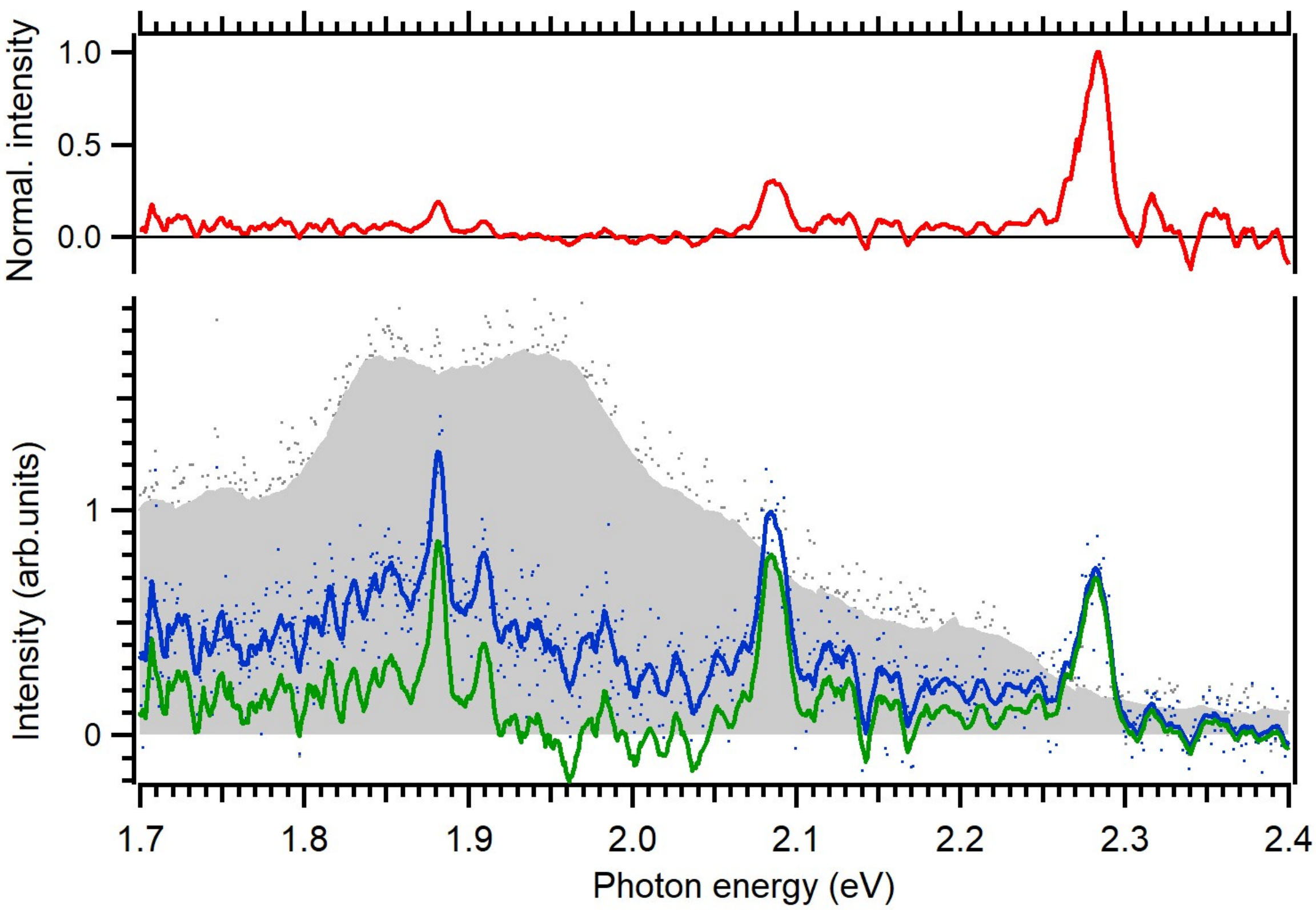

Figure S1. Processing steps of the exciton spectrum shown in Fig. 1c. Points: raw data of the exciton (blue) and of the plasmon (grey) spectrum. Grey area: Savitzky-Golay smoothed plasmon spectrum. Blue line: Savitzky-Golay-smoothed exciton spectrum. Green line: same as blue line but after subtraction of the contribution from the smoothed plasmon spectrum. Red line: fully corrected PdOEP spectrum, same as the green line but after division by the smoothed plasmon spectrum.

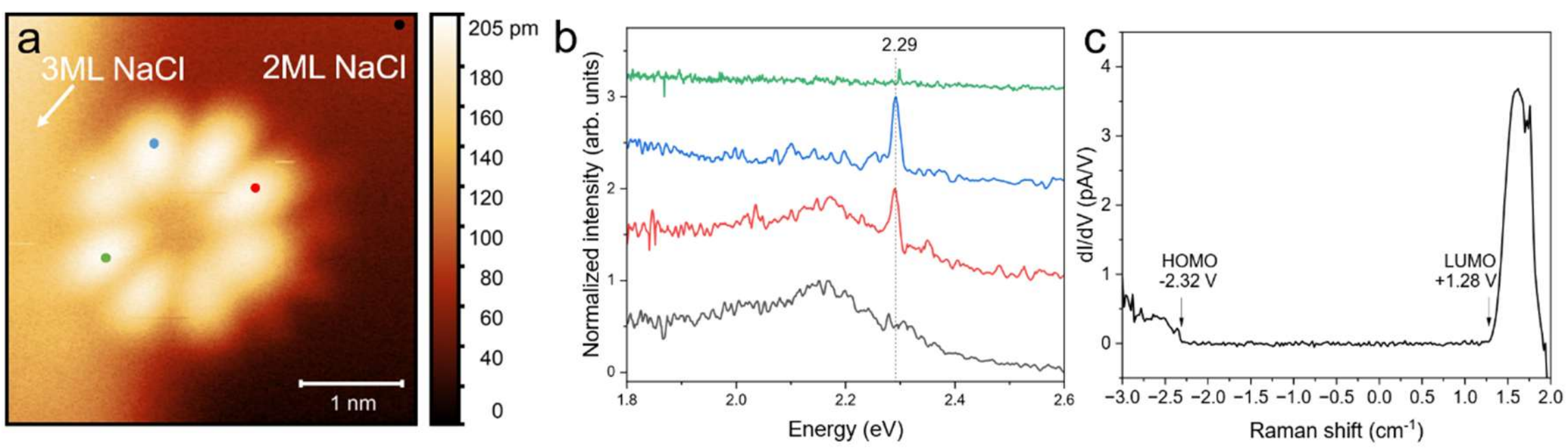

Figure S2. STM image (a) of a PdOEP on 2ML NaCl / Ag(100) and STM electroluminescence spectra (b) of this molecule on different positions. The tip position for each emission measurement is indicated in panel a (tunneling conditions: I = 40 pA, V = −2.5 V, acquisition time is 120 s). The intensities of all the spectra in panel b are normalized to 1 for clarity. The lack of the $T_1$ emission peak can be attributed to the low plasmon enhancement at the energy of the $T_1$ emission and efficient non-radiative $T_1$ quenching channels due to the thinner buffer layer. c. Differential conductance (dI/dV) spectrum of a PdOEP molecule on 2ML NaCl / Ag(100).

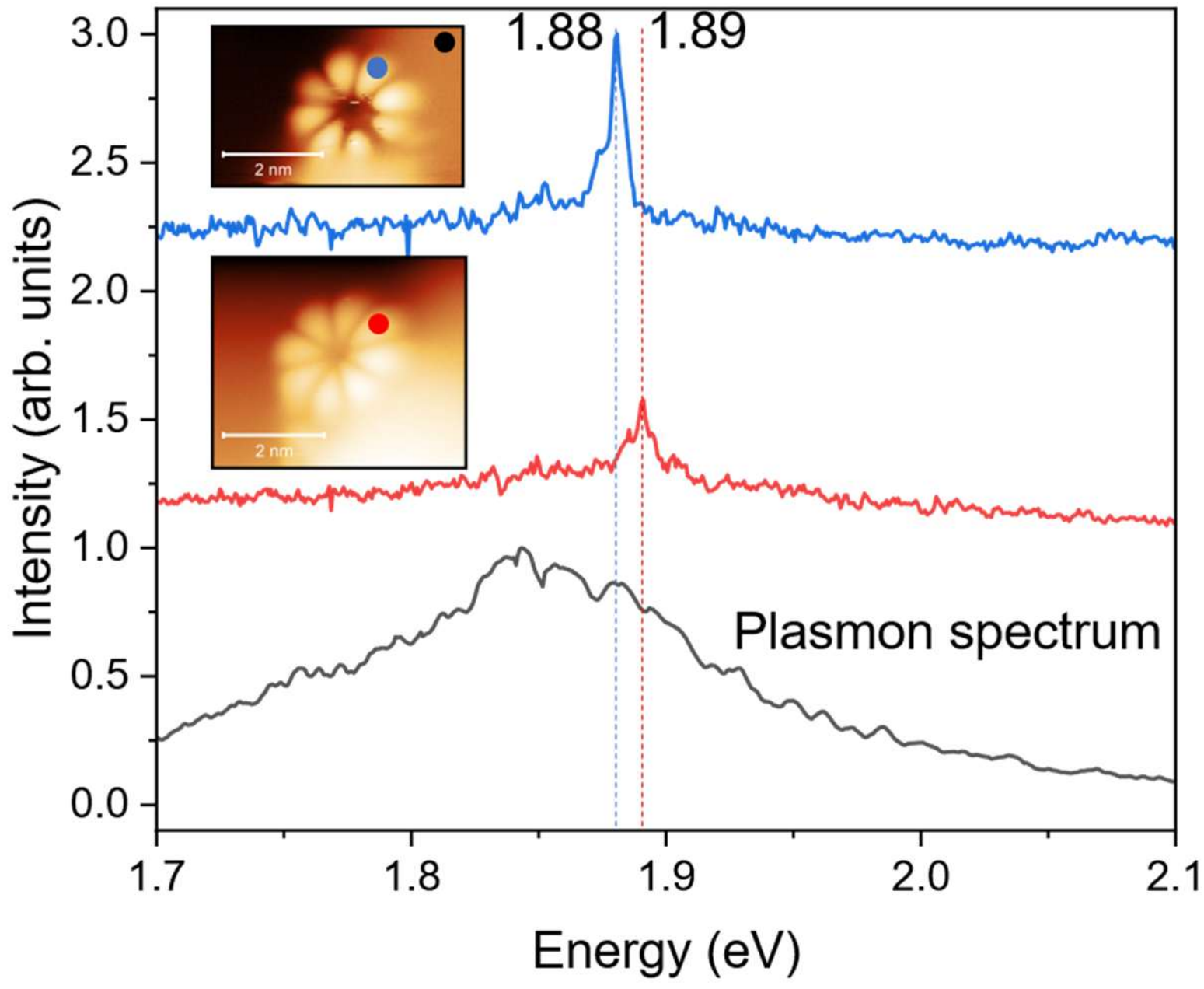

Figure S3. STM electroluminescence spectra (blue and red curves) of the same PdOEP molecule on 3ML NaCl / Ag(100) with different adsorption position of PdOEP on NaCl resulting in a 10 meV shift of the emission energy. The black curve is the plasmon emission spectrum acquired on NaCl near the molecule. Measurement conditions: I = 30 pA, V = −2.5 V, t = 120 s. The blue spectrum is collected on the lobe of the PdOEP molecule when it is adsorbed on the surface of 3ML NaCl, while the red spectrum obtained after this molecule moved and attaches to the edge of a NaCl step, between 2ML and 3ML. The insets show STM images of the molecule in these different adsorption positions with the tip position for the emission measurements indicated. Measurement conditions of the STM images: I = 1 pA, V = −2.5 V, scale bar: 2 nm.

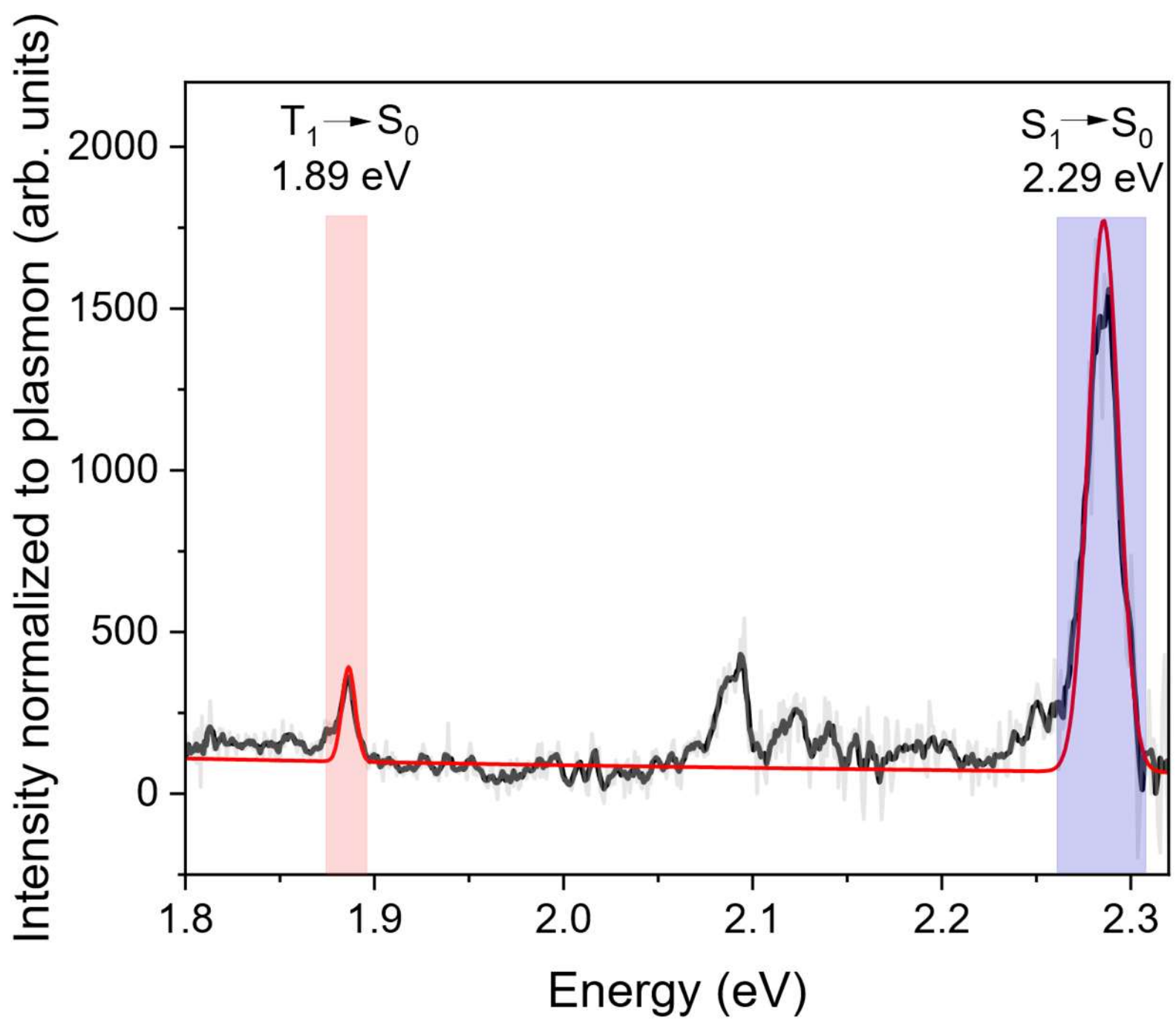

Figure S4. STM electroluminescence spectrum of a PdOEP molecule on 3ML NaCl / Ag(100) with Gaussian fits (red line) of the two emission peaks corresponding to the $S_1$ and $T_1$ emission lines. The spectrum demonstrates a good agreement (less than 4 meV difference) with the peak positions on the Ag(111) substrate (Fig.1 and Fig. 3). Measurement conditions: I = 30 pA, V = −2.5 V, t = 20 s, 150 lines/mm grating. The spectrum is normalized to the plasmonic emission spectrum.

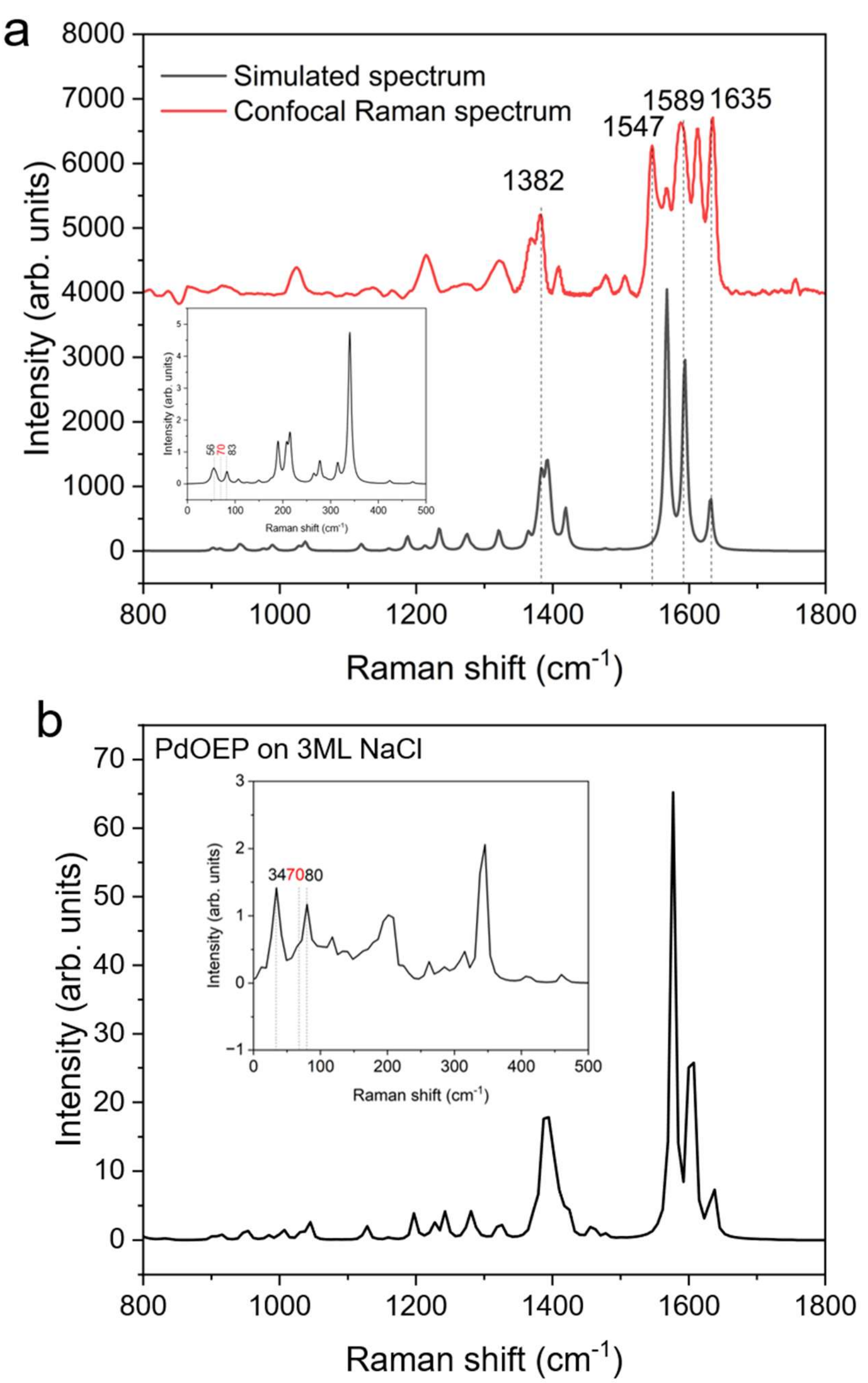

Figure S5. a. Simulated (black curve) and confocal Raman (red curve) spectra of PdOEP. b. Simulated Raman spectrum of PdOEP on 3ML NaCl with Pd on top of Cl. The calculated frequencies were multiplied by a scale factor of 0.95 to correct for the anharmonicity. Insets: Zoomed-in views in the simulated Raman spectra of PdOEP to illustrate the low Raman frequency region. Parameters for the confocal Raman measurement: Excitation laser: 488 nm; Exposure time: 5 min.

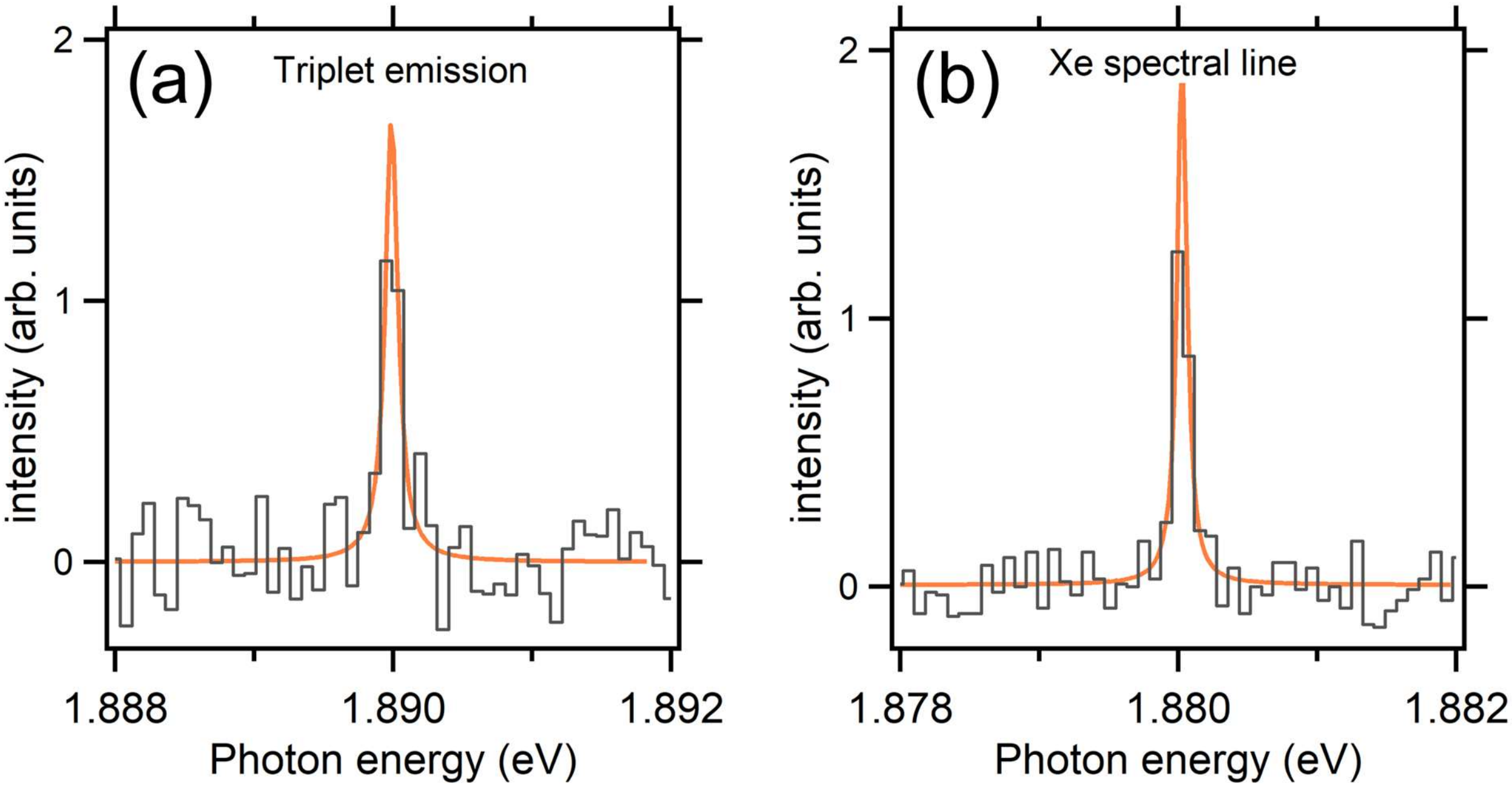

Figure S6. a. spectrum of the triplet emission line shown in Fig.1d; b. energetically close emission line of a Xe spectral lamp recorded with the same set-up using the 1200 lines/mm grating. The comparison between these spectra demonstrates that the width of the triplet emission line is dominated by the instrumental resolution. The orange lines are Lorentzian fits.

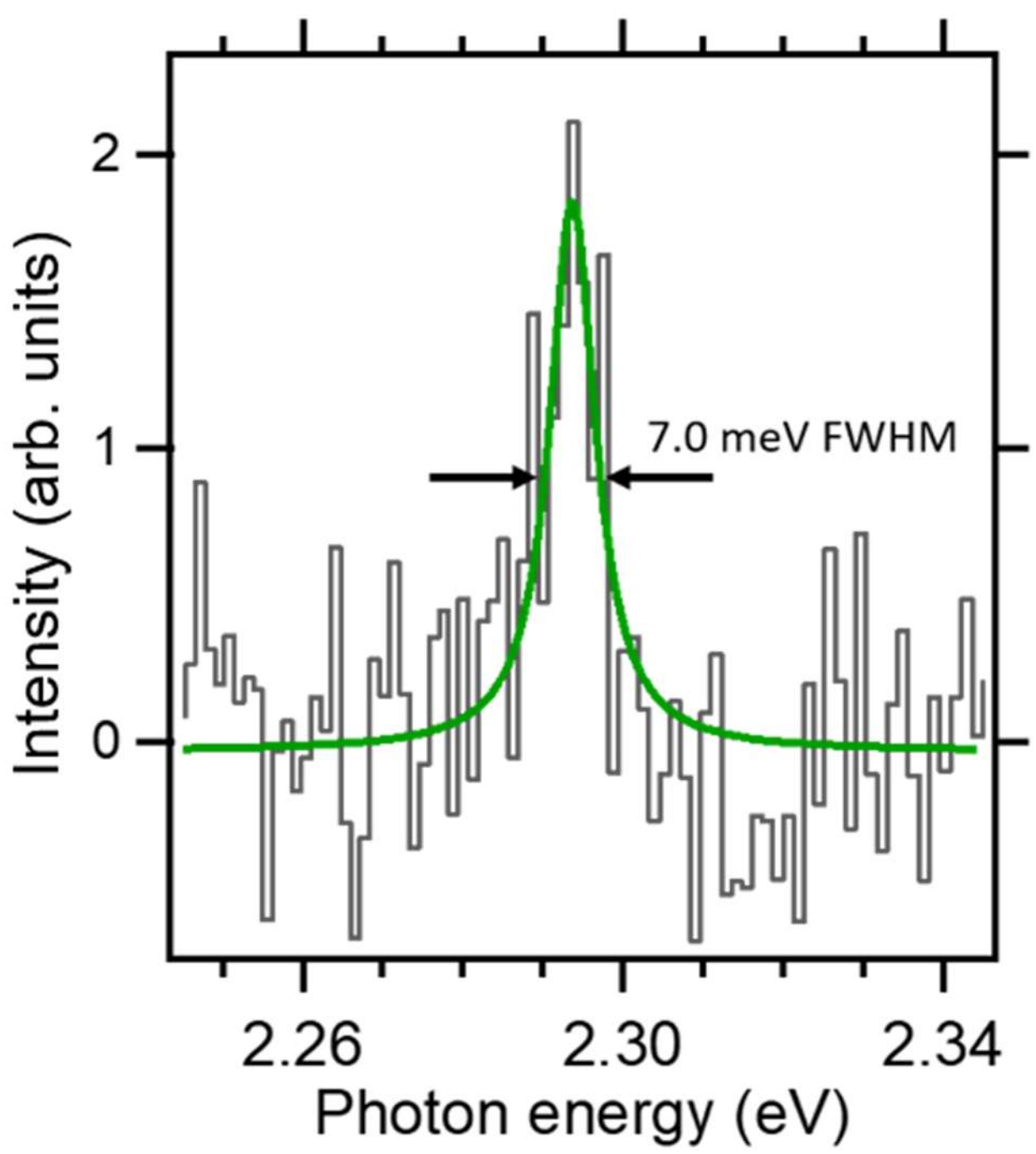

Figure S7. Emission spectrum (black histogram, raw data) of $S_1$ acquired on a PdOEP on 3ML NaCl / Ag(100) with a 1200 line/mm grating. The green line is a Lorentzian fit providing the indicated FWHM of the corresponding 0-0 peak.

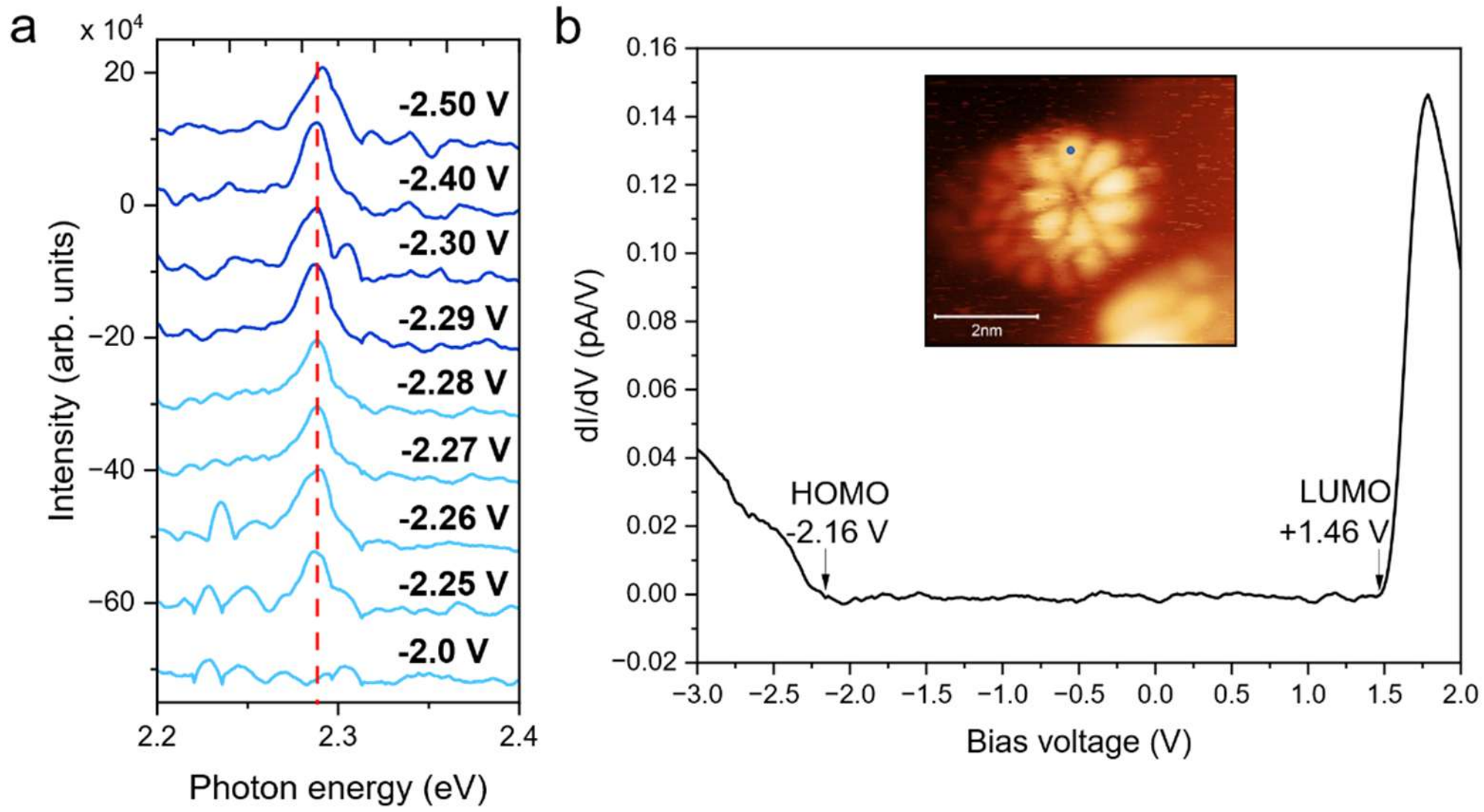

Figure S8. a. STM luminescence spectra of a PdOEP molecule ($S_1$ emission) on 3ML NaCl/Ag(100) as a function of voltage, (I = 30 pA, acquisition time for each spectrum: 120 s). Note that the molecule jumped to the tip at -2.0 V, resulting in the disappearance of the emission peak. The spectra are vertically shifted for clarity. b. Bias spectrum of the PdOEP molecule in panel a. The inset is the STM image of the PdOEP molecules with tip position indicated (blue dot).

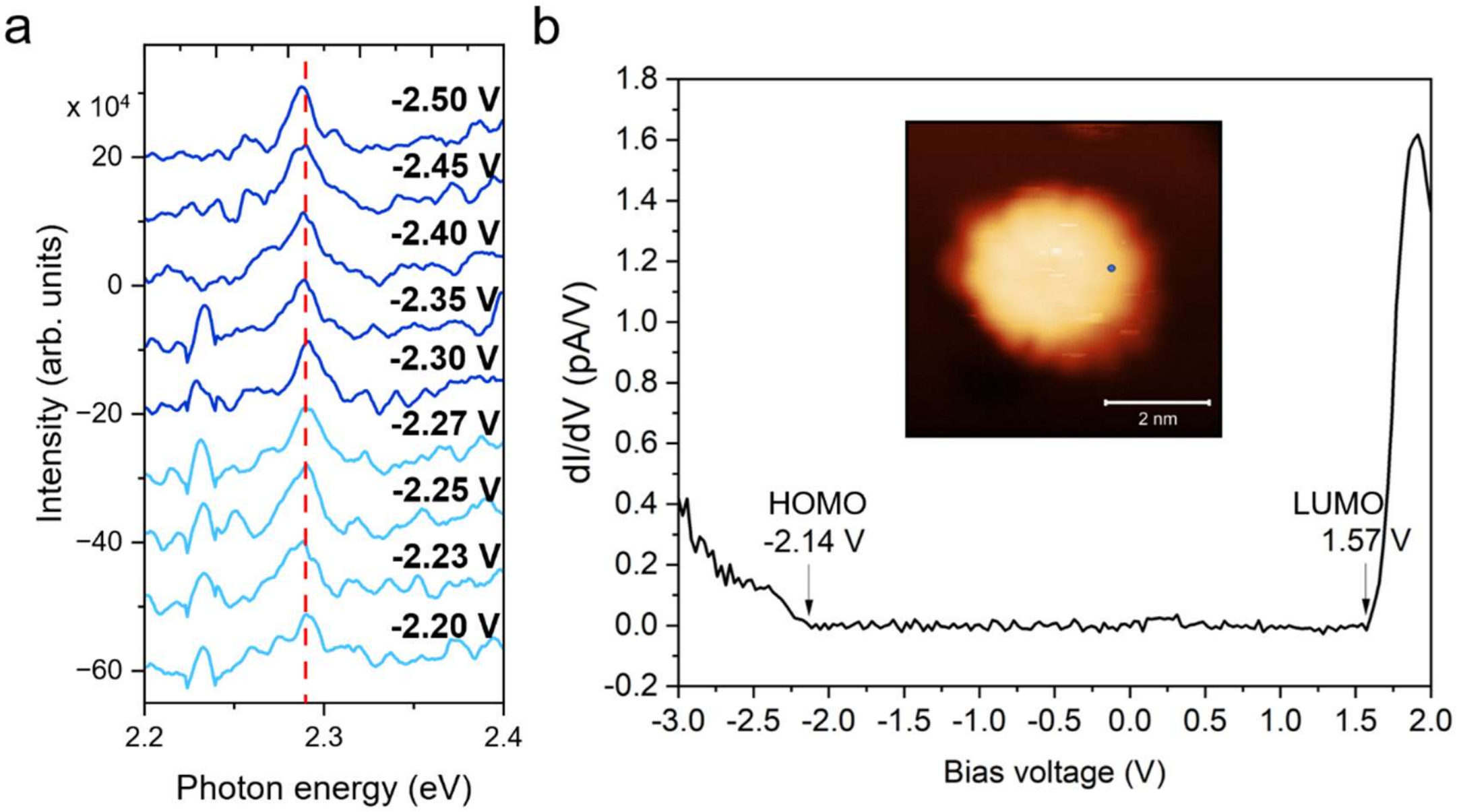

Figure S9. a. STM luminescence spectra of a PdOEP molecule ($S_1$ emission) on 3ML NaCl/Ag(111) as a function of voltage, (**I** = 30 pA, acquisition time for each spectrum: 120 s). The spectra are vertically shifted for clarity. b. Bias spectrum of the PdOEP molecule in panel a. The inset is the STM image of the PdOEP molecule with tip position indicated (blue dot).

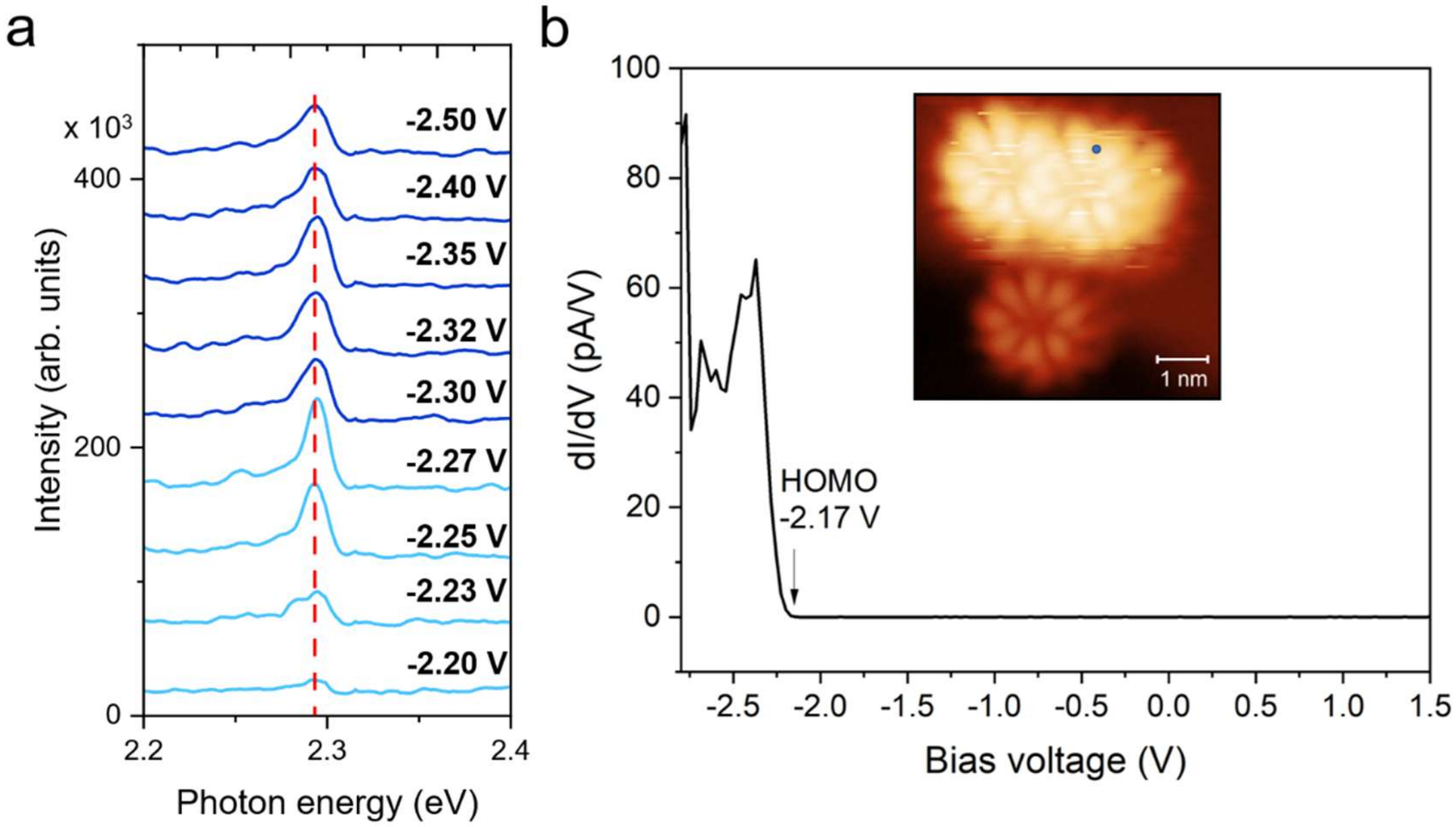

Figure S10. a. STM luminescence spectra of a PdOEP dimer ($S_1$ emission) on 3ML NaCl/Ag(100) as a function of voltage, (I = 30 pA, acquisition time for each spectrum: 120 s). The spectra are vertically shifted for clarity. b. Bias spectrum of the PdOEP molecule in panel a. The inset is the STM image of the PdOEP molecules with tip position indicated (blue dot).

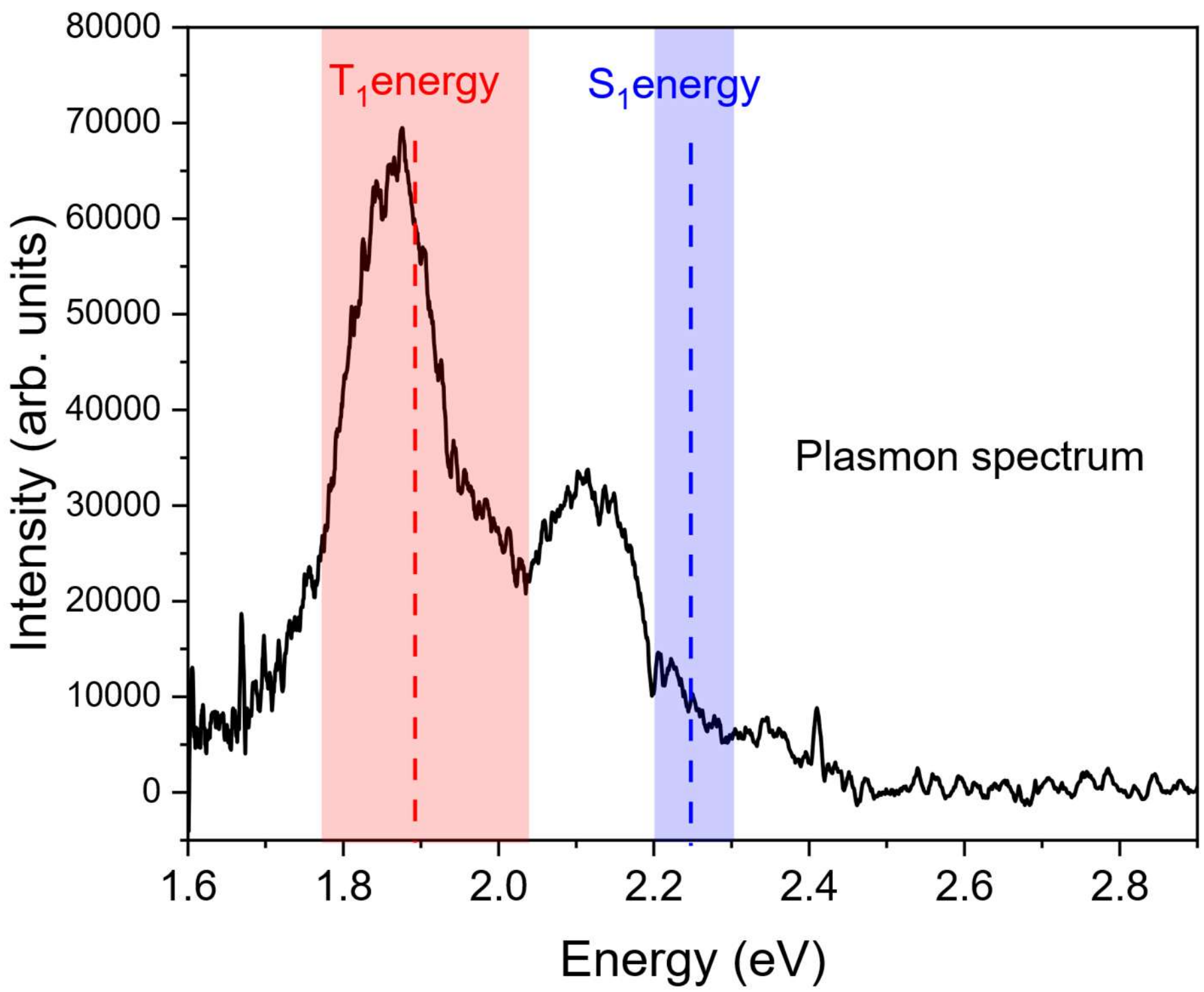

Figure S11. Plasmon spectrum acquired on 3ML NaCl near a PdOEP molecule, whose electroluminescence spectra are shown in Figure 3. The spectrum has been background-subtracted to isolate the plasmonic signal. Measurement conditions: I = 100 pA, V = −2.5 V, t = 150 s.

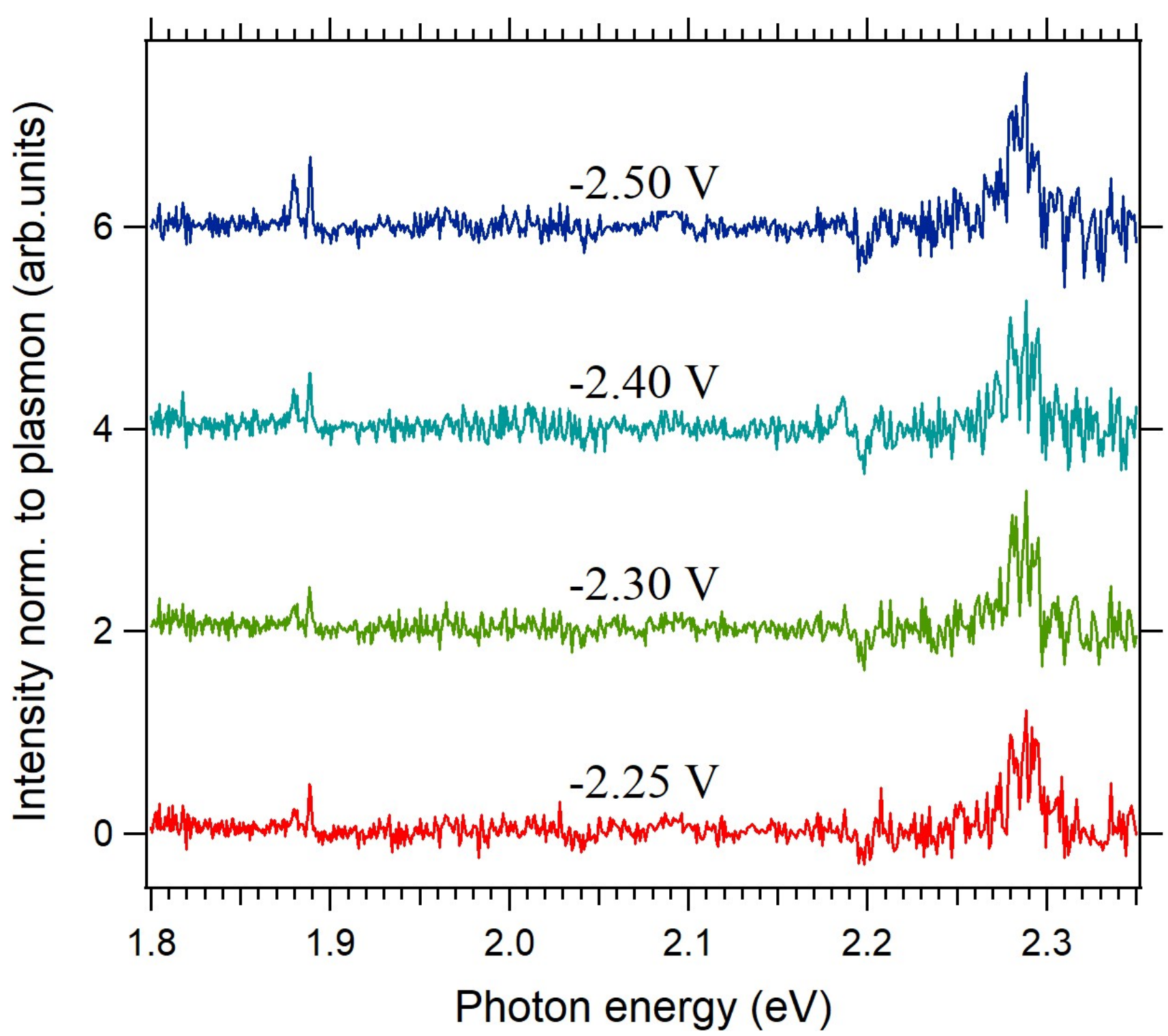

Figure S12. Larger range of the spectra shown in Fig. 3. The spectra are normalized to the relevant plasmon spectrum (Fig. S11). Note that a normalization to the plasmon spectrum induces additional noise. For all spectra the current is I = 42.5 pA, acquisition time per spectrum from top to bottom: 450 s, 600 s, 600 s, and 750 s. The different acquisition times have been compensated in the intensity scale. The spectra are vertically offset for clarity.

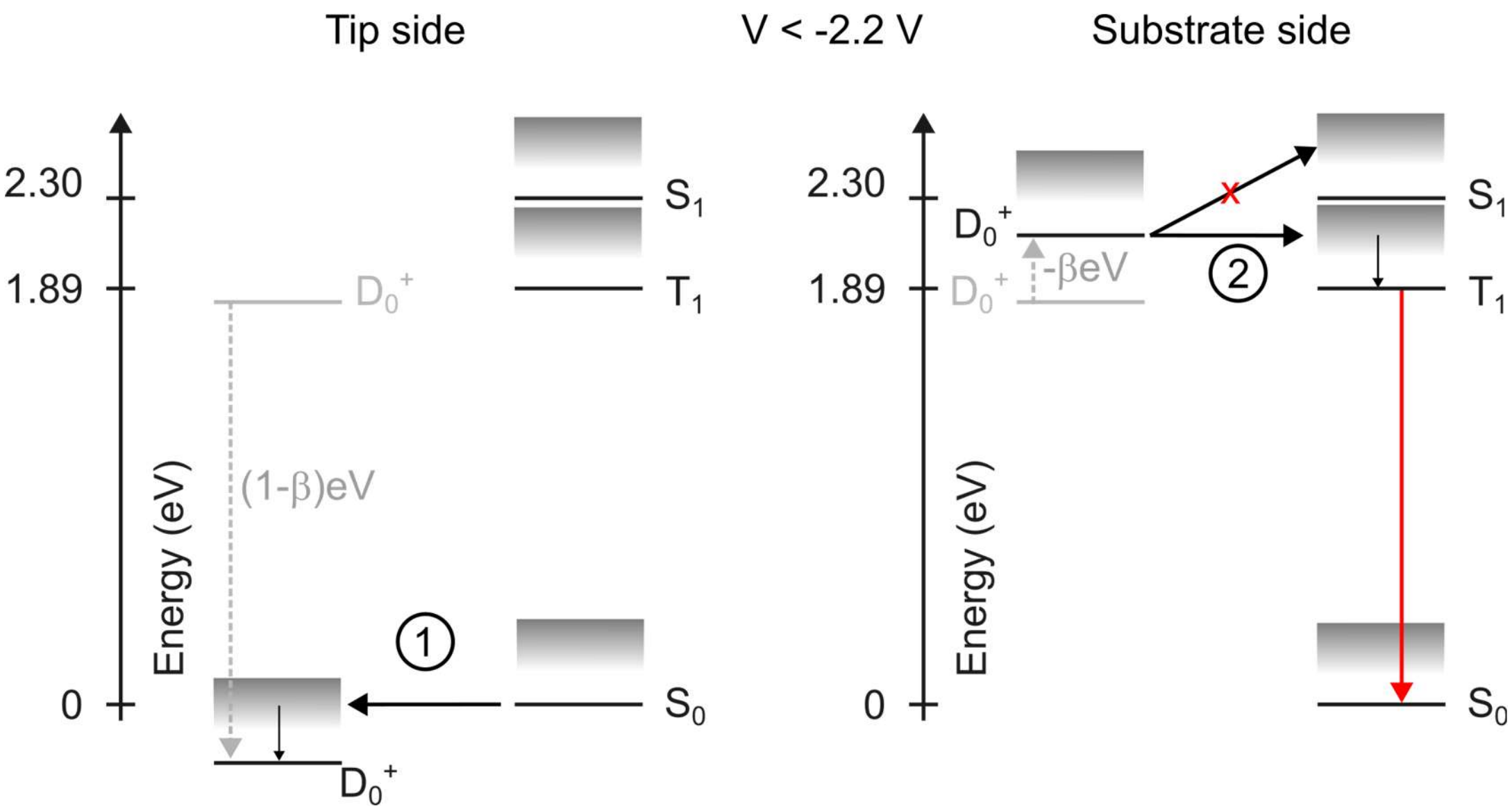

Fig. S13. Many-body diagram including the effects of voltage drop and relaxation energy considering both the tip (left) and the substrate (right) perspective. The gray-shaded areas indicate the relaxation energies, the small vertical arrows illustrate the relaxation processes. We include only 4 states for clarity.

For completeness, in Fig. S13 we discuss the effects of voltage drop and relaxation in the molecule and substrate on the level alignment in the junction, focusing on transitions involving the $D_0^+$ state. Following the work by Kaiser et al.[10], the position of the $D_0^+$ state from the perspective of the substrate can be estimated to be at a (bias-dependent) energy $E_{D0+}(V) = -QV_{res}(1-\beta) - E_{relax} - Q\beta V$. $V_{res}$ is the peak position in the dI/dV (-2.4 V from Fig. 1b), $\beta$ characterizes the voltage drop in NaCl (0.1), $E_{relax}$ is the relaxation energy (estimated to be 0.3 eV), and Q is the total charge of the state measured in e (elementary charge units), which is 1 for $D_0^+$. From the perspective of the tip, the last term is $-Q(1-\beta)V$. These shifts are illustrated in Fig. S9 (left panel), considering $E_{D0+}(0) = 1.86$ eV, which is indicated in gray. On the left, we see that when we consider the transition involving an exchange of electron with the tip, the $D_0^+$ level shifts below $S_0$, so that the tunneling into the continuum reflected in the gray gradient (relaxation energy) is possible (step (1) from the main text). Then, on the substrate side, relevant for the step (2), the voltage shifts the $D_0^+$ state upwards so that its estimated energy is in the 2.08-2.11 eV range for the voltages used in this work. Therefore, the energy of the $D_0^+$ state is not sufficient to excite the $S_1$ (2.29 eV) state directly (arrow marked by "x") and the excitation is driven via the $T_1$ state, as discussed in the main text, justifying the neglect of the effects of the voltage drop and reorganization/relaxation in the NaCl and the molecule.

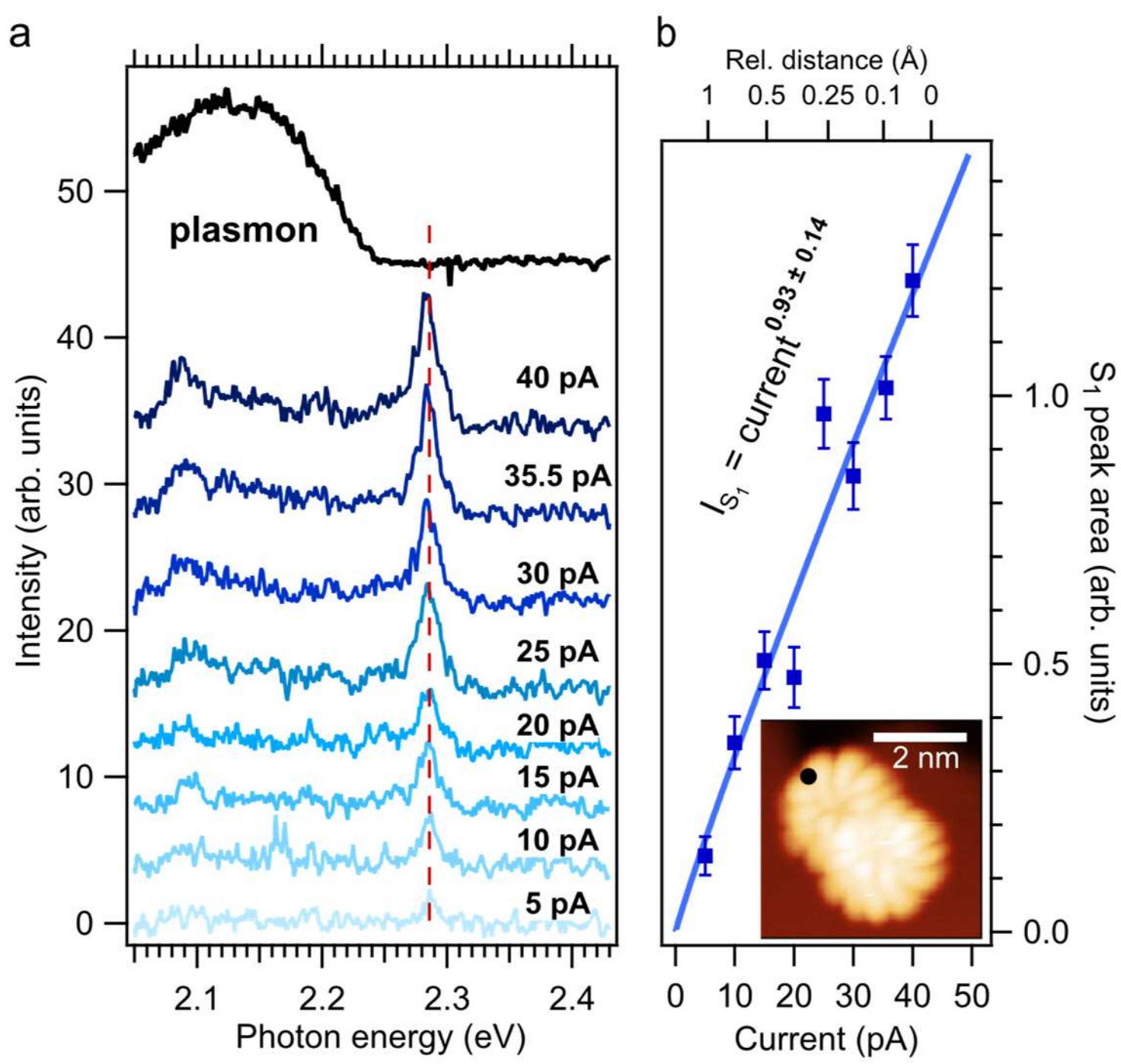

Figure S14. a. Current-dependent STM luminescence spectra of a PdOEP dimer (V = -2.25 V, acquisition time per spectrum: 120 s) and a plasmon spectrum at the same voltage (black curve). b. Fluorescence intensity as a function of current with power law fit. The inset is the STM image of the PdOEP dimer with tip position indicated (black dot). The relative distance scale at the top of the graph indicates the tip height variation for the different current set-points used in the data series.

As an outlook, we investigate the current dependence of the $S_1$ emission of a PdOEP dimer under UCEL conditions. Surprisingly, the $S_1$ emission as a function of current – over the same current range as in Figures 4 and 5 – exhibits a linear or even weakly sub-linear behavior. Assuming that for constant voltage the same injected current leads to a similar molecular dynamic, the result in Figure S16a is incompatible with the close-to-linear increase of the $T_1$ emission. The fit of both curves with one parameter set can only provide very poor fits (see Figure S16b below). In fact, the incompatibility of both current-dependencies finds a straight-forward explanation in the observation that the molecule studied in Fig. S14 has a very close PdOEP neighbor which may allow for a significant deviation from the single molecule up-conversion model from Figure S16a. While we do not obtain a more detailed insight into this interesting two-molecule system, we speculate that the actual mechanism in this case may proceed along a completely different scheme: A first $T_1$ excitation below the tip may hop resonantly to the neighboring molecule so that a second $T_1$ excitation created by the tip also on the first molecule allows a fusion of the two triplets resulting in the formation of an $S_1$ excitation that may eventually emit from either of the two molecules. If the occupation of $T_1$ on the neighboring molecule is sufficiently high and its lifetime is prolonged due to its large distance to the STM tip, the $S_1$ dependence on current may appear to be close to linear, even in the UCEL range. The emission spectra in Fig. S14a exclude that the linearity may be caused by a substantial redshift of the exciton energy due to coupling within the two-molecule system.

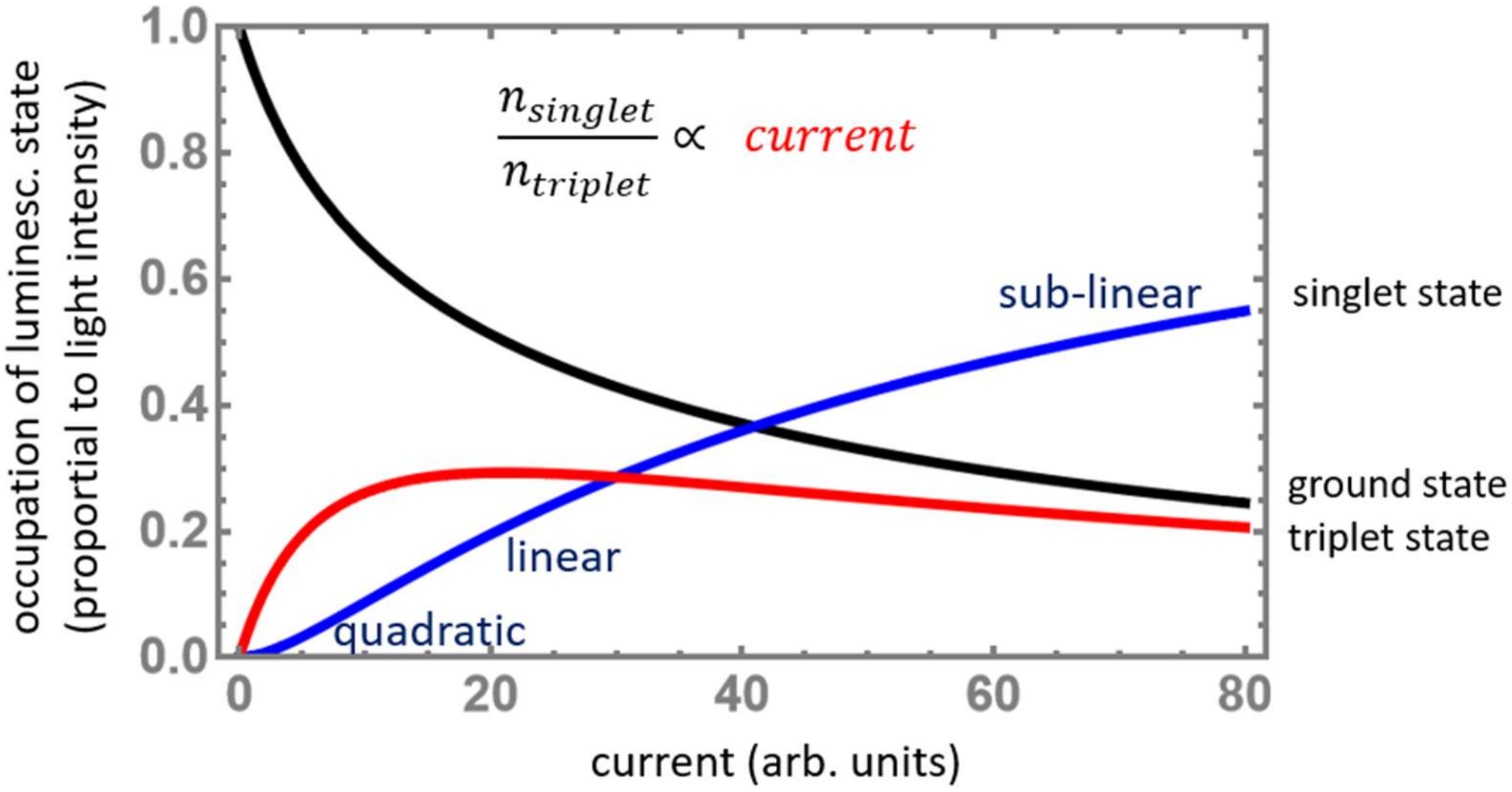

Figure S15. Solution of the up-conversion model for a generic example. Note that the power law behavior of the single state (blue text) changes over a rather narrow range of currents.

Setting up the rate equations (Master equations) for the minimal up-conversion model with the rate constants $a_1$, ... $a_4$, introduced in Fig. S16a, we obtain the following equations for the stationary solution at long times ($t \to \infty$):

$$\frac{dn_0}{dt} = -a_1\, n_0 \;+\; a_3\, n_S \;+\; a_2\, n_T \;\xrightarrow{t\to\infty}\; 0$$

$$\frac{dn_T}{dt} = \quad a_1\, n_0 \;-\; a_2\, n_T \;-\; a_4\, n_T \;\xrightarrow{t\to\infty}\; 0$$

and the conservation of occupation number $n_0 + n_S + n_T \equiv 1$ , or in matrix representation:

$$\begin{pmatrix} a_1 & -a_2 & -a_3 \\ -a_1 & (a_2 + a_4) & 0 \\ 1 & 1 & 1 \end{pmatrix} \begin{pmatrix} n_0 \\ n_T \\ n_S \end{pmatrix} = \begin{pmatrix} 0 \\ 0 \\ 1 \end{pmatrix}$$

In the model, occupation of the singlet state is only possible from the triplet state since direct excitation from the ground state to the singlet state is forbidden by total energy conservation in the UCEL range. Figure S15 sketches a generic solution for the occupation numbers of the ground state ($n_0$), the singlet state ($n_S$) and the triplet state ($n_T$). For simplicity, we disregard occupation of any higher excitonic excitation and the role of intermediate charged states (hence $n_0+n_S+n_T=1$).

The observable emission intensities P are proportional to the average occupation of the emitting state times the rate constant of the emission process. We then obtain the solutions

$$P_{singlet} = \eta_S \, n_s \, a_3 = \eta_S \frac{\boldsymbol{a_1} \, \boldsymbol{a_4} \, a_3}{(a_3 \, (a_2 + \boldsymbol{a_4}) + \boldsymbol{a_1} \, (a_3 + \boldsymbol{a_4}))}$$

$$P_{triplet} = \eta_T \, n_T \, a_2 = \eta_T \frac{\boldsymbol{a_1} \, a_3 \, a_2}{(a_3 \, (a_2 + \boldsymbol{a_4}) + \boldsymbol{a_1} \, (a_3 + \boldsymbol{a_4}))}$$

in which $\eta_S$ and $\eta_T$ are the effective detection efficiencies for the singlet and the triplet transition, respectively. We write the excitation transition rates $a_1$ and $a_4$ with bold symbols in order to indicate that both are proportional to the STM current.

From the solution of the model (Fig. S15) two conclusions can be drawn:

- The ground state depletes continuously as a function of current. The singlet state starts at low currents with a quadratic current dependence, transforming into a linear dependence for intermediate currents and eventually levels off. Over the same current range, the triplet state initially grows linearly, exhibits a plateau and finally decreases again. Depending on the explored current range, the observed power law may thus adopt varying exponents.
- However, we find independent of current the ratio $n_S / n_T = \boldsymbol{a_4} / a_3$ which expresses a proportionality with the current. This means that there is a defined relationship between the current dependencies of singlet and triplet occupation. Since the emission intensities are proportional to the individual occupation numbers (if the emission rates are approximately constant), this relation also holds for the current-dependent emission intensities. We conclude that monitoring the current-dependent intensities can be used to test the up-conversion model even in an experiment without explicit time resolution.

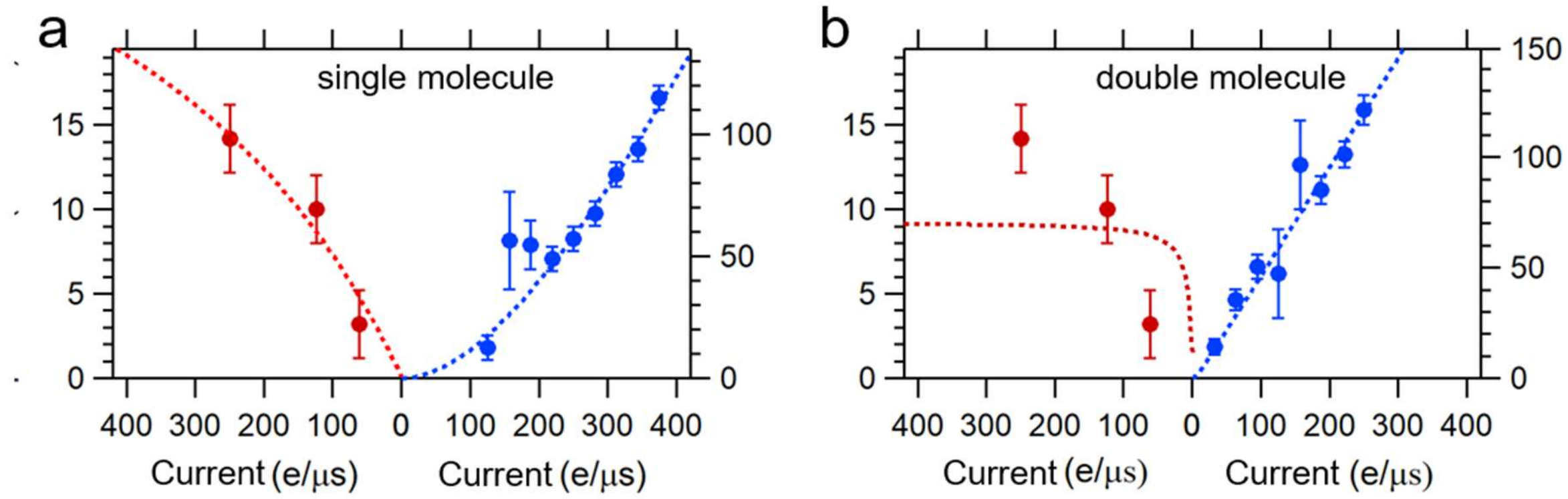

Figure S16. Fit of $S_1$ (blue) and $T_1$ (red) intensities using the parameter sets given in Table S1, (a) for the single molecule case as shown in Fig. 5 and (b) for the case of the double molecule (Fig. S14). Note that the error bars of the outliers in the blue ($S_1$) data have been increased in order to reduce their weight in the fit.

By fitting the observed singlet and triplet emission intensities (respecting their error bars) simultaneously with a single set of fit parameters ($a_1$, .. , $a_4$), we find that absolute rate constants can be obtained only with errors of the same order as the parameter itself. This is due to the fact that the absolute detection efficiencies $\eta$ for $S_1$ and $T_1$ are both unknown and can compensate for large rate constant variations. However, the above result $n_S / n_T \propto current$ imposes remarkably strict conditions on the relation between the current dependencies of both emission intensities reflected in fits of significantly varying fidelity. In Fig. S16 we compare fits for the cases of the single (a) and the double (b) molecule.

For the single molecule we find a qualitatively good description of the data, indicating a slightly sub-linear curve for triplet and a slightly super-linear curve for singlet emission. In the experimentally explored range 10 pA .. 45 pA both curves match well with the model. We thus illustrate that a perfectly linear behaviour of the triplet and a perfectly quadratic behaviour of the singlet are not required by the model, as we already discussed for the generic solution above.

In contrast to the single molecule case (Fig. S16a), the behaviour of the double molecule (Fig. S16b) requires a highly compromised fit that does not describes the triplet behaviour well. As reported in Fig. S14, the singlet behaviour is close to linearity (even slightly sub-linear) and cannot be reconciled with a triplet behaviour which is only slightly sub-linear. Within the employed simple up-conversion model, these two curves are thus incompatible and the linearity of the $S_1$ intensity pushes the $T_1$ intensity fit to become almost independent of current.

**Table S1.** Parameters used for the fits in Figure S15 and the ranges enabling good fits.

| Parameter | Fit Fig.S16a | Ranges for good fits / comments | Fit Fig.S16b |
|---|---|---|---|
| $a_1$ | $0.5 * I/e$ | $0.1 * I/e < a_1 < I/e$ | $0.5 * I/e$ |
| $a_2$ | $0.48*10^9$ $s^{-1}$ | $0.5 *10^9$ $s^{-1} < a_2 < 10^{15}$ $s^{-1}$ | $6.0*10^6$ $s^{-1}$ |
| $a_3$ | $10^{11}$ $s^{-1}$ | $10^6$ $s^{-1} < a_3 < 10^{15}$ $s^{-1}$ | $10^{11}$ $s^{-1}$ |
| $a_4$ | $0.5 * I/e$ | $0.1 * I/e < a_4 < I/e$ | $0.5 * I/e$ |
| $\eta_S$ | $2.7*10^{-6}$ | scaling factor well defined by fitting | $2.0*10^{-6}$ |
| $\eta_T$ | $1.8*10^{-7}$ | scaling factor well defined by fitting | $3.1*10^{-6}$ |

We find that each parameter can be separately varied over several orders of magnitude while optimizing the other parameters still provides a good fit. The only sharp limit found is the lower bound of $a_2$ (marked in red), which indicates that no reasonable fit can be obtained when the rate constant $a_2$ is set smaller than $0.5*10^9$ $s^{-1}$, thus suggesting a time constant below 2 ns in good agreement with the range (5 ps – 1 ns) estimated in the main text in the discussion of Figure 6. In the table, *I* denotes the tunnel current and *e* denotes the elementary charge.

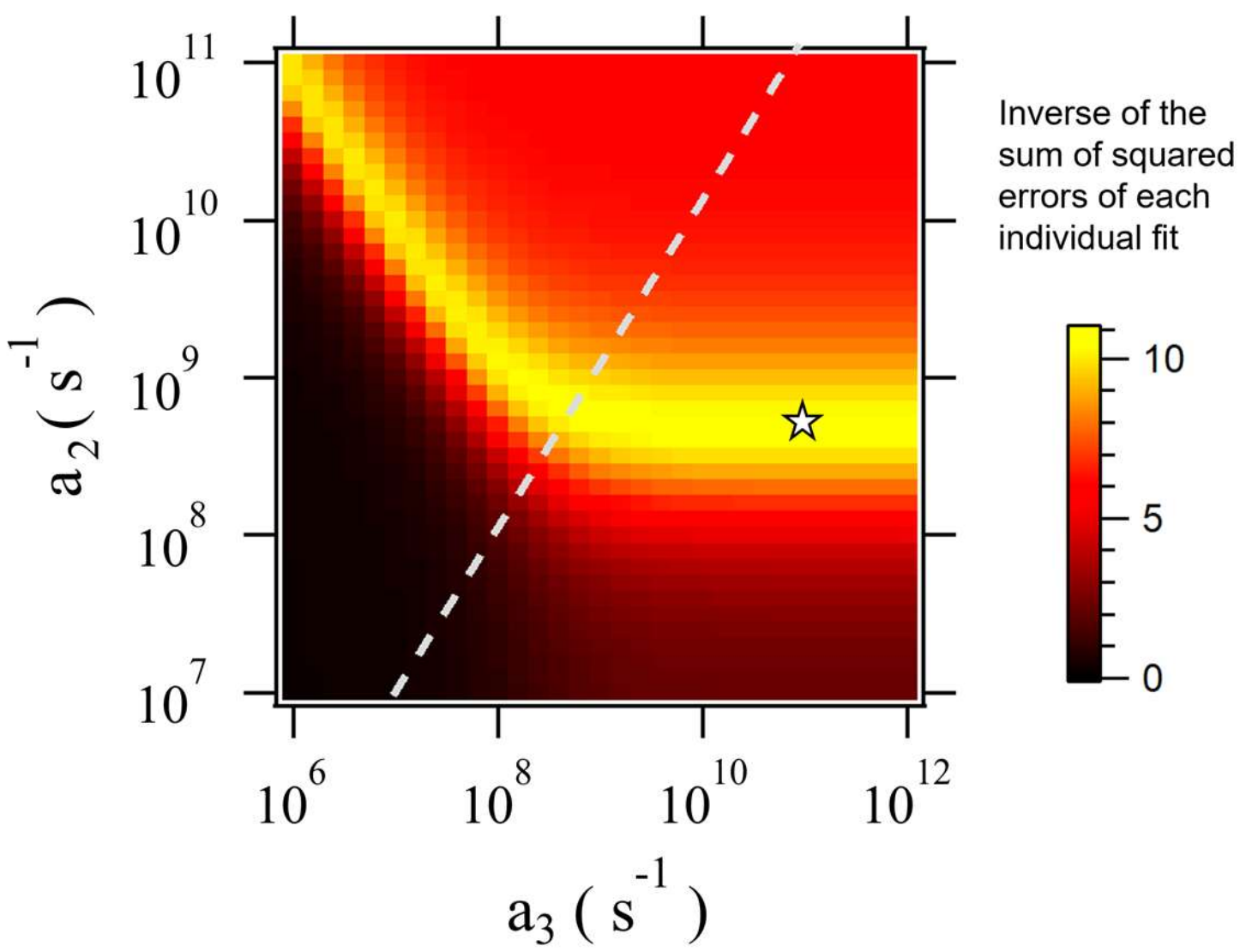

Figure S17. Plot of fit quality for the data of Figure S16a, illustrating the relation between the two decay rate constants $a_2$ and $a_3$. As in Table S1, $a_1$ and $a_4$ are assumed to be 0.5 * *I/e*. In this plot, only the data scaling parameters $\eta_S$ and $\eta_T$ are fitted. Yellow areas indicate good, black areas very poor fitting results. The plot illustrates the existence of an absolute $a_2$ limit discussed in Table caption S1. The star marks the parameter combination in Fig. S16a. If we demand that the triplet decay rate is smaller than the one of the singlet, only combinations on the right hand side of the white dashed line shall be considered.